\documentclass[a4paper,12pt]{article}

\usepackage{mathrsfs}
\usepackage{graphicx}
\usepackage{bm}
\usepackage{amssymb}
\usepackage{amsmath}
\usepackage{ascmac}
\usepackage{array}

\usepackage{cite}

\begin{document}



\title{Second law, entropy production, and reversibility in thermodynamics of information}
\author{Takahiro Sagawa
 \footnote{\scriptsize Department of Applied Physics, The University of Tokyo, 7-3-1 Hongo, Bunkyo-ku, Tokyo 113-8656, Japan}}

\date{}

\maketitle

\begin{abstract}
We present a pedagogical review of the fundamental concepts in thermodynamics of information,  by  focusing on the second law of thermodynamics and the entropy production.
Especially, we discuss the relationship among thermodynamic reversibility,  logical reversibility, and heat emission in the context of the Landauer principle and clarify that these three concepts are fundamentally distinct to each other.
We also discuss thermodynamics of measurement and feedback control by Maxwell's demon.  
We clarify that the demon and the second law   are indeed consistent in the measurement and the feedback processes individually, by including the mutual information to the entropy production.
\end{abstract}

\tableofcontents


\section{Introduction}
\label{Sec_Introduction}

Thermodynamics is intrinsically related to information, as the entropy represents the lack of our knowledge.  
The first clue on the information-thermodynamics link  was provided by Maxwell, who considered a thought experiment of ``Maxwell's demon''~\cite{Maxwell}.  
Later, Szilard suggested a quantitative connection between work extraction and information gain~\cite{Szilard}, even several decades before the establishment of information theory by Shannon~\cite{Shannon}.
The role of information in thermodynamics was investigated and controversies were raised throughout the twentieth century~\cite{Demon,Brillouin,Landauer,Bennett,Zurek1989,Landauer2,Goto}.

In this decade, thermodynamics of information has attracted renewed attention because of the development of the modern theory of nonequilibrium thermodynamics~\cite{Cohen1993,Gallavotti,Evans2002,Jarzynski1997,Gaveau1997,Crooks1999,Jarzynski2000,Seifert2005,Kawai2007,Esposito2011,Sekimoto,Jarzynski2011,Seifert2012}, which is often referred to as stochastic thermodynamics. Especially, a fundamental thermodynamic relation called  the fluctuation theorem was discovered in 1990s~\cite{Cohen1993,Gallavotti,Jarzynski1997,Crooks1999}, which has opened up a new avenue of research.  
We note, however, that only a few seminal works have been done already in 1970s and 80s~\cite{Kawasaki,Schnakenberg,Bochkov,Broeck1986,Mou1986}.

Thermodynamics of information can now be formulated  based on  stochastic thermodynamics~\cite{Parrondo2015}, by incorporating information concepts such as Shannon entropy and mutual information.
Specifically, the Landauer principle for information erasure~\cite{Shizume1995,Piechocinska2000,Barkeshli2005,Lutz2009,Turgut2009,Sagawa-Ueda2009,Reeb2014,Sagawa2014} and feedback control by Maxwell's demon~\cite{Touchette2000,Touchette2004,Kim2007,Sagawa-Ueda2008,Cao2009,Sagawa-Ueda2010,Sagawa2012a,Horowitz2010,Sagawa-Ueda2012a,Abreu2012,Lahiri2012,Still2012,Sagawa-Ueda2012b,Sagawa-Ueda2013} have been investigated from the modern point of view.

Furthermore, the relationship between thermodynamics and information has become significant from the experimental point of view~\cite{Ciliberto2017}.
The first quantitative demonstration of the work extraction by Maxwell's demon was performed in Ref.~\cite{Toyabe2010}, and  the Landauer bound for information erasure was demonstrated in Ref.~\cite{Berut2012}.
Several fundamental experiments have been further performed  both in the classical and quantum regimes~\cite{Rodan2014,Koski2014,Koski2015,Chida2017,Berut2013,Jun2014,Gavrilov2016a,Gavrilov2017,Hong2016,Lopez-Suarez2016,Silva2016,Vidrighin2016,Camati2016,Cottet2017,Masuyama2017}.

In this article, we review the theoretical foundation of thermodynamics of information.
Especially, we aim at clarifying the concept of reversibilities and the consistency between Maxwell's and the second law, which we hope would unravel subtle conceptual problems.

We will put special emphasis on the following two observations. First, reversibility has several different aspects.  In particular, thermodynamic reversibility and logical reversibility are fundamentally distinct concepts, which are associated with different kinds of the degrees of freedom of a thermodynamic system.  Second, mutual information is a crucial concept to understand the consistency between the demon and the second law.  We will see that the demon is consistent with the second law for the measurement and the feedback processes \textit{individually}.

We here make some side remarks.  First, in this article, we only consider classical systems in the presence of an infinitely large heat bath, though essentially the same argument applies to quantum systems.
Second, this article is completely newly written, but is closely related to a paper~\cite{Sagawa2014} by the author, where the present article is intended to be more pedagogical and comprehensive.
Finally, for simplicity of notation, we set the Boltzmann constant to unity (i.e., $k_{\rm B}=1$) throughout the article.

This article is organized as follows.  In the rest of this section, we briefly summarize the above-mentioned two observations.
In Sec.~\ref{Sec_Reversibility_conventional}, we review the second law and reversibility in conventional thermodynamics as a preliminary.
In Sec.~\ref{Sec_Reversibility_stochastic}, we discuss the framework of stochastic thermodynamics and clarify what reversibility means there.
In Sec.~\ref{Sec_Reversibility_computation}, we review reversibility in computation, which is referred to as logical reversibility.
In  Sec.~\ref{Sec_Landauer}, we discuss thermodynamics of information in a simple setup, and state the Landauer principle.  The relationship between thermodynamic reversibility and logical reversibility is clarified in this simple setup.
In Sec.~\ref{Sec_Thermodynamics_computation}, we generally formulate thermodynamics of computation.
In Sec.~\ref{Sec_Work}, we slightly change the topic and discuss work extraction by Maxwell's demon.  In particular, we consider the upper bound of extractable work and formulate thermodynamic reversibility with feedback.
In Sec.~\ref{Sec_Entropy}, we generally discuss the entropy balance during the measurement and the feedback processes of the demon and clarify how the demon is consistent with the second law in these processes.
In Sec.~\ref{Sec_Concluding}, we make concluding remarks, where we briefly summarize some topics that are not mentioned in the preceding sections.

\

$\Diamond \Diamond \Diamond$

\

\begin{figure}[htbp]
 \begin{center}
 \includegraphics[width=100mm]{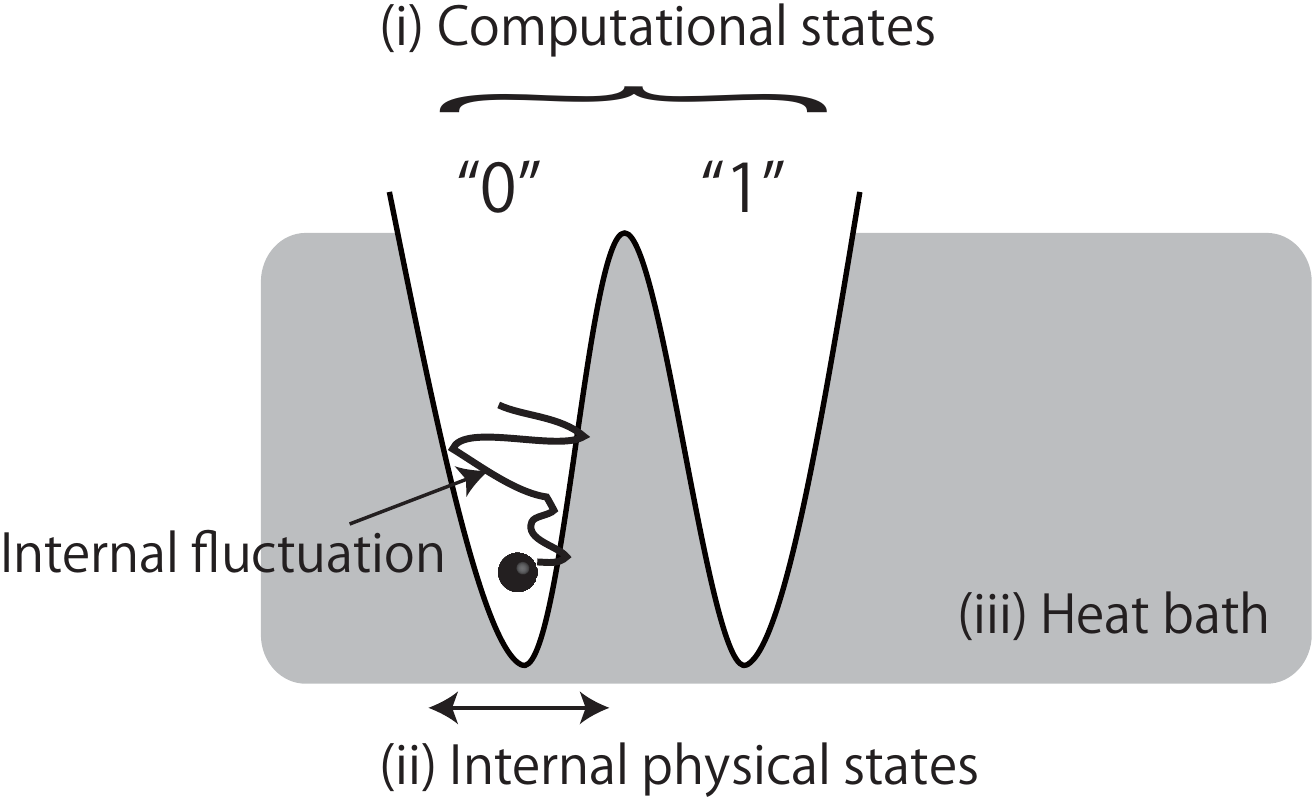}
 \end{center}
 \caption{A schematic of  a binary memory, which is modeled as a Brownian particle in a double-well potential.  The memory represents computational state ``$0$'' or ``$1$'', when the particle is in the left or right well, respectively.  These wells are separated by a barrier that is sufficiently higher than thermal fluctuations.  The internal physical degrees of freedom represent the position of the particle inside individual wells, where the particle suffers thermal fluctuations inside these wells.  The entire memory is attached to a heat bath that is in thermal equilibrium.} 
\label{Intro_fig}
\end{figure}

At this stage, let us summarize some key observations, which will be detailed in the following sections. We consider a binary memory that stores one bit of information (``$0$'' and ``$1$''), and suppose that the memory is in contact with a  single heat bath (see Fig.~\ref{Intro_fig}).  
We focus on the relationship among  thermodynamic reversibility,  logical reversibility, and heat emission from the memory.
We first note that there are three kinds of the degrees of freedom in this setup: 
\begin{description}
\item[(i)] The computational states  of the memory (i.e.,  ``$0$'' and ``$1$'' for the binary case).  Such computational states should be robust against thermal fluctuations, in order to store information stably. 
\item[(ii)] Internal physical states of the memory, which represent the physical degrees of freedom associated with a single computational state.  
\item[(iii)] The degrees of freedom of the heat bath, which is assumed to be in thermal equilibrium.
\end{description}
Then, we have the following observations:
\begin{itemize}
\item Thermodynamic reversibility refers to the reversibility of the total system including the heat bath and thus is connected to the entropy change in (i)+(ii)+(iii).
\item Logical reversibility refers to the reversibility of the computational states only and thus is connected to the entropy change in  (i).
\item Heat transfer to the bath is bounded by the  entropy  change of all the degrees of freedom of the memory, i.e.,  (i)+(ii).
\end{itemize}
Therefore, the above three concepts should be distinguished fundamentally, while some of them can become equivalent in some specific setups.

Entropy production characterizes  thermodynamic reversibility, and thus is related to the degrees of freedom (i)+(ii)+(iii).  A general version of the second law of thermodynamics states that entropy production is always nonnegative for any transition  from a nonequilibrium state to another nonequilibrium state (see Sec.~\ref{Sec_Reversibility_stochastic} for details).
In particular, the entropy production of the total system, including an engine and the memory of Maxwell's demon, is nonnegative for individual processes of measurement and feedback control.  A crucial point here is that the mutual information between the engine and the demon should be counted as a part of the entropy production.  By doing so, the demon is always consistent with the second law of thermodynamics,  and we do not need to consider the information-erasure process to understand the consistency.

\section{Reversibility in conventional thermodynamics}
\label{Sec_Reversibility_conventional}

As a preliminary, we briefly review conventional thermodynamics, which has been established in the nineteenth century as  a phenomenological theory for macroscopic systems.
A remarkable feature of conventional thermodynamics lies in the fact that it can be formulated in a self-contained manner, without referring to underlying microscopic dynamics such as Newtonian mechanics and quantum mechanics.
In fact, an equilibrium state is characterized by only a few macroscopic quantities such as the energy and the temperature.
We can also define the thermodynamic entropy in a purely phenomenological manner by using, for example, the Clausius formula~\cite{Callen}. 

We note that conventional thermodynamics can be formulated as a mathematically rigorous axiomatic theory~\cite{Lieb1999}.
While in this article we do not formalize  our argument in a rigorous manner,  the following argument of this section can be made rigorous in line with the theory of Lieb and Yngvason~\cite{Lieb1999}.

We focus on the situation that a thermodynamic system is in contact with a single heat bath at temperature  $T$.
Let $\beta := T^{-1}$ be the inverse temperature.
We consider a transition from an equilibrium state to another equilibrium state.
During the transition, the system absorbs the heat $Q$ from the bath, and changes its thermodynamic entropy by $\Delta S_{\rm T}$.
We note that  in conventional thermodynamics, the thermodynamic entropy $S_{\rm T}$ is defined only for equilibrium states, and the second law only concerns a transition from an equilibrium state to another equilibrium state, though intermediate states can be out of equilibrium.

Then, the second law is stated as follows.\\
\\
\textbf{Second law of conventional thermodynamics.} \textit{An equilibrium state can be converted into another equilibrium state with heat absorption Q, if and only if}
\begin{equation}
\Delta S_{\rm T} - \beta Q \geq 0.
\label{second_law1}
\end{equation}

\

We note that the ``only if'' part (i.e., any possible state conversion satisfies inequality (\ref{second_law1})) is the usual second law, while ``if'' part (i.e., a state conversion is possible if  inequality (\ref{second_law1}) is satisfied) is also true under reasonable axioms~\cite{Lieb1999}.

The left-hand side of the second law~(\ref{second_law1}) is referred to as the entropy production, which we denote by
\begin{equation}
\Sigma := \Delta S_{\rm T} - \beta Q.
\end{equation}
This terminology, the entropy production, dates back to Prigogine, who associated $-\beta Q$ with the entropy change of the bath~\cite{Prigogine1998}.  In this spirit, $\Sigma$ is regarded as the entropy change of the entire ``universe'' that consists of the system and the heat bath.

We can also rewrite the second law (\ref{second_law1}) in terms of the work and the free energy.  Let $W$ be the work performed on the system, and $\Delta E$ be the change in the average internal energy.  The first law of thermodynamics is given by
\begin{equation}
W+Q = \Delta E.
\label{first_law0}
\end{equation}
By substituting the first law into inequality~(\ref{second_law1}), we obtain
\begin{equation}
W \geq \Delta F_{\rm eq},
\label{eq_free_second}
\end{equation}
where 
\begin{equation}
F_{\rm eq} := E - TS_{\rm T}
\end{equation}
is the equilibrium free energy.  If the process is cyclic, inequality~(\ref{eq_free_second}) reduces to $W \geq 0$, which is the Kelvin's principle, stating that perpetual motion of the second kind is impossible (i.e., a positive amount of work cannot be extracted from an isothermal cycle).

\

$\Diamond\Diamond\Diamond$

\

We next formulate the concept of reversibility in conventional thermodynamics.  Based on the standard textbook argument~\cite{Callen}, we adopt the following definition:\\
\\
\textbf{Definition (Reversibility in conventional thermodynamics).} {\it A state transition from one to another equilibrium state is thermodynamically reversible, if and only if the final state can be restored to the initial state, without remaining any  effect on the outside world.}

\

We note that ``effect'' above is regarded as a ``macroscopic effect'' in conventional thermodynamics because microscopic changes (i.e., the subleading terms in the thermodynamic limit) are usually neglected.

A crucial feature of this definition is that thermodynamic reversibility is completely characterized by the entropy production, as represented by the following theorem.\\
\\
\textbf{Theorem.} {\it Thermodynamic reversibility is achieved if and only if the entropy production is zero, i.e.,} $\Sigma = 0$.\\
\\
\textbf{Proof.} While one can find a proof of the above theorem in standard textbooks of thermodynamics (at least implicitly), we reproduce it here for the sake of self-containedness.

i)  Suppose that a transition is thermodynamically reversible.  Then there exists a reverse transition that satisfies the requirements in the definition of reversibility.  From the requirement that there is no remaining effect in the outside world, $Q_{\rm reverse} = -Q$ should hold, because otherwise the energy of the heat bath is changed after the reverse transition.  Combining this with $\Delta S_{\rm T, reverse} = - \Delta S_{\rm T}$, we have $\Sigma_{\rm reverse} = -\Sigma$.  On the other hand, both of $\Sigma \geq 0$ and $\Sigma_{\rm reverse} \geq 0$ hold from of the second law.  Therefore, $\Sigma = \Sigma_{\rm reverse} = 0$.

ii) Suppose that $\Sigma$ is zero.  Then $- \Sigma = (-\Delta S_{\rm T}) - \beta (-Q)$ is also zero.  Therefore, the reverse transition is possible with $Q_{\rm reverse} := - Q$, because of the ``if'' part of the second law.\hspace{\fill}$\Box$

\

We consider the concept of quasi-static process, by which we define that the system remains in equilibrium during the entire process.
Thermodynamic reversibility is achieved if a process is quasi-static, because the quasi-static condition guarantees that $\Delta S_{\rm T} \simeq \beta Q$.
The quasi-static limit is achieved by an infinitely slow process in many situations in which
\begin{equation}
\Delta S_{\rm T} =\beta Q + O(\tau^{-1})
\end{equation}
holds, where $\tau$ is the time interval of the entire process.  In the infinitely-slow limit $\tau \to +\infty$, the equality in (\ref{second_law1}) is achieved.
More precisely, $\tau \to + \infty$ means $\tau / \tau_0 \to \infty$, where $\tau_0$ is the relaxation time of the system.

We note, however, that there is a subtle difference between quasi-static and infinitely slow processes in general. 
For example, suppose that there is a box separated by a wall at the center, and gas is only in the left side of the wall.
No matter how slow the removal of the wall is, the free expansion of the gas to the right side is not quasi-static and thermodynamically irreversible.
Such a situation has been experimentally demonstrated in Ref.~\cite{Gavrilov2016b} in the context of stochastic thermodynamics.

The foregoing argument can be straightforwardly generalized to situations with multiple heat baths at different temperatures.
In particular, the zero entropy production of a heat engine with two baths implies the maximum efficiency of the heat-to-work conversion.
For example, the Carnot cycle attains the maximum efficiency, where the entropy production is zero and the cycle is thermodynamically reversible.
We note that it has been rigorously proved that any thermodynamically reversible process is infinitely slow, based on  some reasonable assumptions (including that fluctuations of the system do not diverge)~\cite{Shiraishi2016a}.
Since an infinitely slow process gives the zero power (i.e., the work per unit time is zero), thermodynamically reversible engines might be practically useless.  
For that reason, the efficiency at the maximum power has been intensively studied~\cite{
Curzon1975,Esposito2010}.

The concept of reversibility in conventional thermodynamics is generalized to stochastic thermodynamics, as discussed in the next section.

\section{Reversibility in stochastic thermodynamics}
\label{Sec_Reversibility_stochastic}

Stochastic thermodynamics is an extension of thermodynamics to situations where a system is not necessarily macroscopic, and the initial and final states are not necessarily in thermal equilibrium~\cite{Sekimoto,Jarzynski2011,Seifert2012}.
When  a large heat bath is attached to a small system, thermal fluctuations affect  the system, and its dynamics become stochastic.
Correspondingly, thermodynamic quantities, such as the heat, the work, and the entropy, become stochastic.
Biochemical molecular motors and colloidal particles are typical examples of stochastic thermodynamic systems, with which numerous experiments have been performed~\cite{Ciliberto2017}.

Because of thermal fluctuations, the second law of thermodynamics can be violated  with a small probability in small systems.
At the level of the ensemble average, however, the second law is still valid in an extended form.
Furthermore, a universal relation called the fluctuation theorem has been established by taking into account the role of thermal fluctuations of the entropy production, from which the second law of thermodynamics can be reproduced.
This is the reason why thermodynamics is still relevant to small systems.

To formulate the second law of  stochastic thermodynamics, we need the concept of information entropy, in particular Shannon entropy.
Let $X$ be a probability variable, which takes a particular value $x$ with probability $P(x)$.  The Shannon entropy of $X$ is then defined as~\cite{Cover-Thomas}
\begin{equation}
S(X) := -\sum_{x} P(x) \ln P(x) \geq 0.
\end{equation}
If the probability variable is continuous, we replace the summation above by an integral over $x$.
We note that if  $P(x) = 0$, $P(x) \ln P(x)$ is regarded as zero.

In contrast to the thermodynamic entropy that can be defined only for thermal equilibrium, the Shannon entropy can be defined for an arbitrary probability distribution.  However, these entropies coincide in the canonical distribution $P_{\rm can}(x) := e^{\beta (F_{\rm eq}-E(x))}$, where $F_{\rm eq}$ is the equilibrium free energy and $E(x)$ is the Hamiltonian (i.e., the internal energy of state $x$).
In this case, the Shannon entropy is given by the difference between the average energy and the free energy:
\begin{equation}
S(X) := -\sum_{x} P_{\rm can}(x) \ln P_{\rm can}(x) =  \beta \left( \sum_x P_{\rm can}(x) E(x) - F_{\rm eq} \right),
\label{entropy_free_energy}
\end{equation}
which is a statistical-mechanical expression of the thermodynamic entropy.

We suppose that a small system is described by a Markov jump process or  overdamped Langevin dynamics~\cite{Kampen,Gardiner}.  We also suppose that the system is driven by external parameters (e.g., the center and the frequency of optical tweezers), which are represented by time-dependent parameters in a master equation or a Langevin equation.
The following argument is independent of the details of  dynamics, and therefore we do not explicitly write down stochastic equations.

On the other hand, we assume that variables that break the time-reversal symmetry (i.e., that have the odd parity for the time-reversal transformation) are absent or negligible.  In particular, the momentum term is negligible in dynamics of the system (in particular, the Langevin equation is overdamped), and the magnetic field is absent in the external potential.  
  While some of the following arguments are straightforwardly generalized to systems with the momentum and the magnetic field,  there are some subtle problems  with odd-parity variables.

Let $x$ be the state of the system, which has a certain probability distribution.
Correspondingly, we define the Shannon entropy of the system, written as $S$.
We consider a transition from the initial distribution to the final distribution of the system, where both the distributions are arbitrary and can be out of equilibrium.
Let $\Delta S$ be the change in the Shannon entropy of the system, and $Q$ be the ensemble average of the heat absorption.
We then have the following version of the second law:\\
\\
\textbf{Second law of stochastic thermodynamics.} \textit{A distribution can be converted into another distribution with the heat absorption Q, if and only if}
\begin{equation}
\Delta S - \beta Q \geq 0.
\label{second_law2}
\end{equation}

\

As in conventional thermodynamics, the left-hand side of inequality~(\ref{second_law2}) is referred to as the  (ensemble averaged) entropy production:
\begin{equation}
\Sigma := \Delta S - \beta Q.
\label{entropy_production2}
\end{equation}
This is regarded as the total entropy increase in the ``whole universe'' that consists of the system and the heat bath.

As discussed before, if the distribution is canonical, the Shannon entropy reduces to the thermodynamic entropy.  Therefore, inequality~(\ref{second_law2}) is a reasonable generalization of the conventional second law~(\ref{second_law1}) to situations that the initial and final distributions are not necessarily canonical.

We can rewrite inequality (\ref{second_law2}) in terms of the work and the free energy.  Let $W$ be the work performed on the system. 
We denote the average energy of the system by $E :=\sum_x P(x) E(x)$.
The first law of thermodynamics is again given by 
\begin{equation}
W+Q= \Delta E,
\label{first_law}
\end{equation}
where $\Delta E$ is the change in the average energy.  By substituting (\ref{first_law}) into inequality (\ref{second_law2}), we obtain
\begin{equation}
W \geq \Delta F,
\label{second_law_work2}
\end{equation}
where 
\begin{equation}
F := E - T S
\label{noneq_free}
\end{equation}
is called the nonequilibrium free energy~\cite{Gaveau1997,Esposito2011}.
We note that if the probability distribution is canonical, we have from Eq.~(\ref{entropy_free_energy}) that $F = F_{\rm eq}$.  In such a case, inequality (\ref{second_law_work2}) reduces to $W \geq \Delta F_{\rm eq}$, which is nothing but the second law of equilibrium thermodynamics.

We now consider the precise  meaning of  ``if and only if'' in the second law above.
The ``only if'' part (i.e., any possible transition satisfies inequality (\ref{second_law2}))   has been proved  for Markovian stochastic dynamics, including Markov jump processes and Langevin dynamics~\cite{Gaveau1997,Crooks1999,Seifert2005,Esposito2011}.  Furthermore,  inequality  (\ref{second_law2}) has been proved for a setup in which the total system, including the heat bath, obeys Hamiltonian dynamics~\cite{Jarzynski2000}.

\begin{figure}[htbp]
 \begin{center}
 \includegraphics[width=100mm]{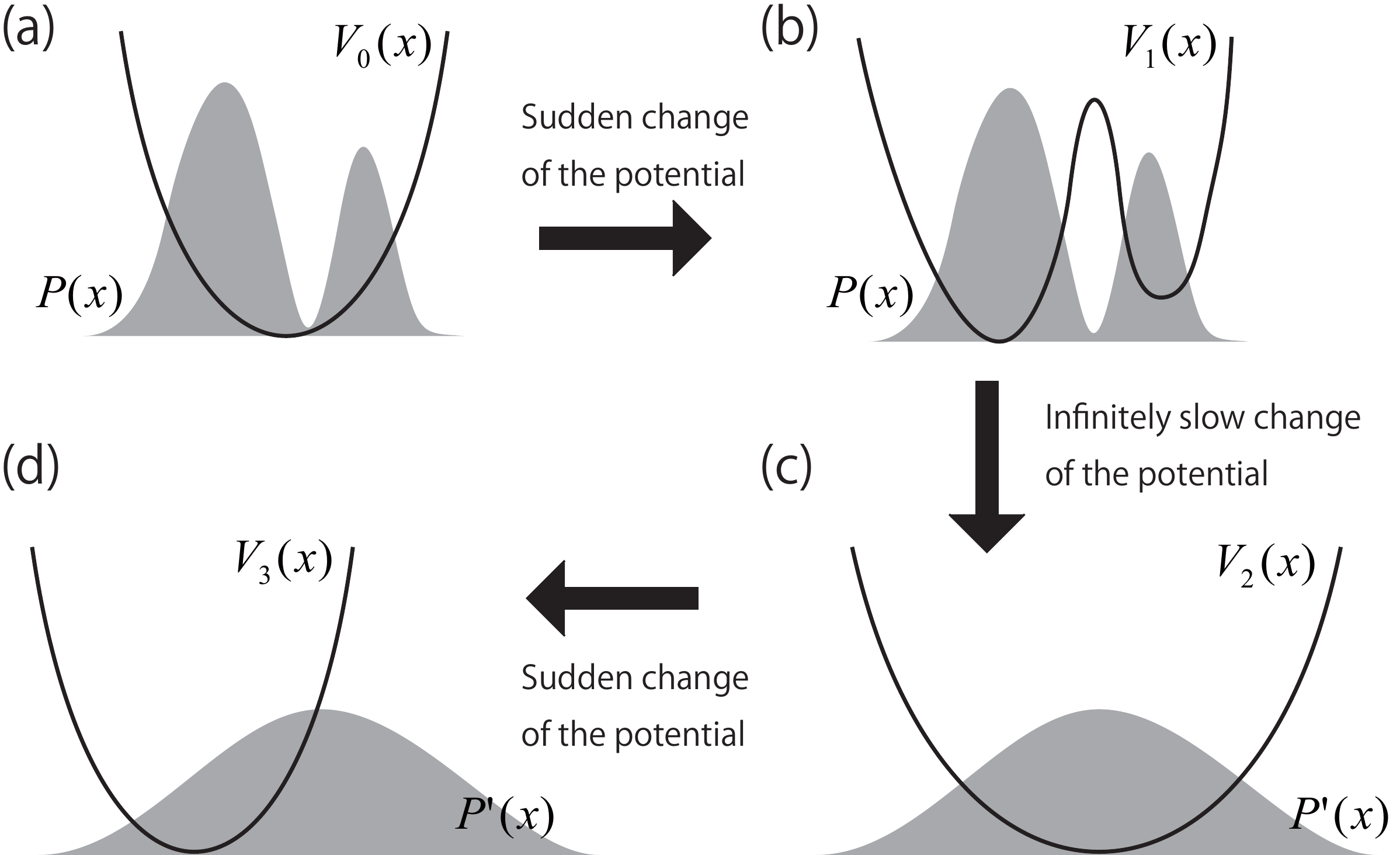}
 \end{center}
 \caption{Protocol that achieves thermodynamic reversibility~\cite{Parrondo2015}.  (a) The probability distribution $P(x)$ (shaded) is in general different from the canonical distribution of the external potential $V_0(x)$ (i.e., the system is out of equilibrium).  (b) The instantaneous change of the potential from $V_0(x)$ to $V_1 (x)$ such that $P(x)$ is now the canonical distribution of $V_1(x)$. (c) The potential is infinitely slowly changed from $V_1(x)$ to $V_2(x)$. Correspondingly, the distribution changes infinitely slowly, and ends up with $P'(x)$ that is the canonical distribution of $V_2 (x)$. (d) The potential is again suddenly changed from $V_2(x)$ to $V_3(x)$.  The distribution $P'(x)$ is in general no longer the canonical distribution of $V_3(x)$. 
During the entire dynamics, the probability distribution does not evolve spontaneously, which makes the entropy production zero.}
\label{Reversible_protocol}
\end{figure}

Before discussing the ``if'' part, we show that there exists a  protocol that achieves the equality in (\ref{second_law2}), for given initial and final distributions and potentials~\cite{Parrondo2015}.
Such a protocol is schematically shown in Fig.~\ref{Reversible_protocol}, which consists of sudden and infinitely slow changes of the external potential.  
While the initial distribution can be out of equilibrium, the potential is instantaneously adjusted to the distribution to make it equilibrium.
Then, the potential is changed infinitely slowly, which brings the distribution to the final one.
After that, the potential is again changed suddenly, and the final distribution can again be out of equilibrium.

It is easy to check that the entropy production is zero,   $\Sigma = 0$, in this protocol.
In fact, the heat absorption from (b) to (c) satisfies $\Delta S - \beta Q = 0$, while in the potential switching  processes the heat absorption is zero and the Shannon-entropy change is also zero.    Therefore, the entropy production of the entire process is zero.
The essence of this protocol is that the distribution is always canonical, except for the very moments of the initial and final times.
In this sense, this protocol is regarded as quasi-static.
In other words, the probability distribution never spontaneously evolves during the entire process, which prohibits the entropy production from becoming positive.

Based on the above protocol, the ``if'' part of the  second law  (i.e., a transition is possible if  inequality (\ref{second_law2}) is satisfied)  can be shown, by explicitly constructing a protocol with $\Delta S - \beta Q > 0$.
For example, we can add  an auxiliary cyclic process to an intermediate step of the infinitely slow protocol ((b)-(c) in Fig.~\ref{Reversible_protocol}). 
If such a cyclic process is not slow, it simply  ``stirs'' the system, and the system emits a positive amount of heat (i.e., $-Q' > 0$).
By adding this heat emission, we have a positive amount of  entropy production in the entire process.

\

$\Diamond\Diamond\Diamond$

\

We next discuss the concept of reversibility in stochastic thermodynamics.
While the fundamental idea is the same as in conventional thermodynamics,  we need to care about probability distributions in stochastic thermodynamics.  We thus adopt the following definition:\\
\\
\textbf{Definition (Reversibility in stochastic thermodynamics).} {\it A stochastic process is thermodynamically reversible, if and only if the final probability distribution can be restored to the initial one, without remaining any  effect on the outside world.}

\

As in conventional thermodynamics, reversibility defined above is completely characterized by the entropy production.\\
\\
\textbf{Theorem.} {\it Reversibility in stochastic thermodynamics is achieved if and only if the entropy production is zero, i.e.,} $\Sigma = 0$.

\

The proof  of this theorem is completely parallel to the case of conventional thermodynamics, just by replacing thermodynamic entropy by Shannon entropy.

From the above theorem,  the protocol described in Fig.~\ref{Reversible_protocol}, satisfying $\Sigma = 0$, is thermodynamically reversible.  We can also directly see this, because the final distribution is restored to the initial distribution, just by reversing the entire protocol step by step.  In this reversed protocol, the heat absorption satisfies $Q_{\rm reverse} = - Q$, and therefore no effect remains in the outside world.

\section{Reversibility in computation}
\label{Sec_Reversibility_computation}

We next discuss the concept of reversibility in computation, which we will show is fundamentally distinct from thermodynamic reversibility.

Let $M$ be the set of the input states of computation.  For example, if any input consists of $n$ binary bits, then $M = \{ 0,1 \}^n$.  We can also consider $M'$ being the set of  the output states of computation, which can be different from $M$ in general.  
Any computation process is a map $\hat C$ from $M$ to $M'$.

We see three simple examples of such computation.

\begin{description}
\item[NOT] The NOT gate simply flips the input bit: $M=M'= \{ 0,1 \}$, and 
\begin{equation}
\hat C(0) = 1, \ \hat C(1) = 0.
\end{equation}
\item[ERASE] The information erasure maps any input to a single ``standard state.''  If the input is one bit and the standard state is ``$0$'', then    $M=M'= \{ 0,1 \}$, and 
\begin{equation}
\hat C(0) = 0, \ \hat C(1) = 0.
\end{equation}
\item[AND] For the AND gate,  the input is two bits and the output is one bit: $M = \{ 0,1 \}^2$, $M' = \{ 0,1 \}$, and
\begin{equation}
\hat C(00) = 0, \ \hat C(01) = 0, \ \hat C(10) = 0, \ \hat C(11) = 1.
\end{equation}
\end{description}

Rigorously speaking, a {\it  computable} map from $\mathbb N^k$ to $\mathbb N$ is defined as a partial recursive function, or equivalently, a partial function that can be implemented by a Turing machine~\cite{Moore2011}.  However, this precise characterization of computability is not necessary for the following argument.  We also note that we only consider deterministic computation in this article.

We now define  logical reversibility of computation~\cite{Landauer,Bennett,Landauer2}.  For a logically reversible computation, one can recover the original input from  only the output, which is formalized as follows.\\
\\
\textbf{Definition (Logical reversibility).} {\it A deterministic computational process $\hat C$ is logically reversible, if and only if it is an injection.
In other words, $\hat C$ is logically reversible if and only if, for any output, there is a unique input.}

\

\begin{figure}[htbp]
 \begin{center}
 \includegraphics[width=70mm]{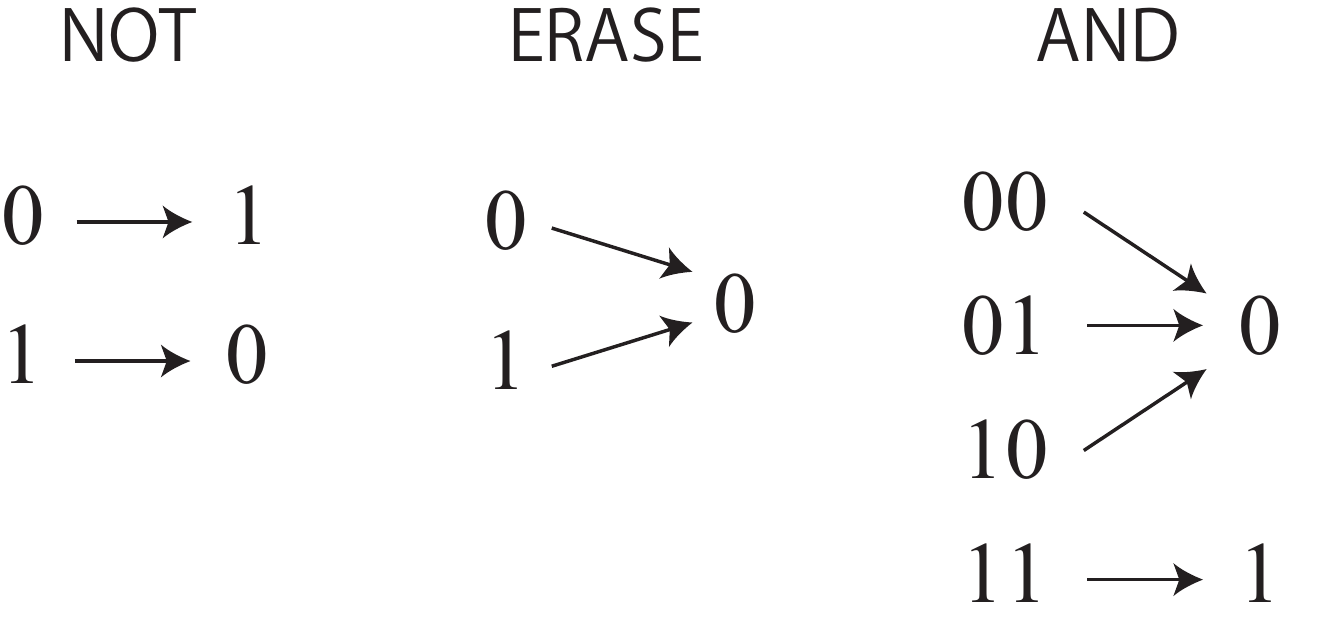}
 \end{center}
 \caption{Three examples of computation. NOT is logically reversible, while ERASE and AND are logically irreversible.}  
\label{computation} 
\end{figure}

In the case of the aforementioned three examples, NOT is logically reversible, while ERASE and AND are logically irreversible.
Figure \ref{computation}  schematically illustrates these examples, where it is visually obvious that only NOT is injection and thus logically reversible.

We next show that logical reversibility can be characterized by the Shannon entropy of the computational states.
For that purpose, we consider a probability distribution over inputs.  Let $p(m)$ be the probability of  input $m \in M$.  The probability distribution over the outputs is then given by
\begin{equation}
p(m') = \sum_{m  \in \hat C^{-1} (m')} p(m),
\label{pm_sum}
\end{equation}
where $m  \in \hat C^{-1} (m')$ means  $m' = \hat C (m)$.
Correspondingly, we define the Shannon entropies of the input and the output by
\begin{equation}
S(M) := -\sum_{m \in M} p(m) \ln p(m), \  S(M') := -\sum_{m' \in M'} p(m') \ln p(m').
\end{equation}

Then, as a general property of the Shannon entropy~\cite{Cover-Thomas}, we have
\begin{equation}
\Delta S(M) := S(M')- S(M) \leq 0.
\label{S_monotone}
\end{equation}
In fact,
\begin{equation}
\begin{split}
S(M)- S(M') &=- \sum_{m' \in M'}  \sum_{m  \in \hat C^{-1} (m')}  p(m) \ln p(m) + \sum_{m' \in M'}  \sum_{m  \in \hat C^{-1} (m')} p(m) \ln p(m')  \\
&= \sum_{m' \in M'}  \sum_{m  \in \hat C^{-1} (m')} p(m) \ln \frac{p(m')}{p(m)}\\
&\geq 0,
\end{split}
\label{S_monotone_proof}
\end{equation}
where we used Eq.~(\ref{pm_sum}) to obtain the second term on the right-hand side of the first line, and used $p(m') \geq p(m)$ with $m' = \hat C (m)$ to obtain the last inequality.
Therefore, the Shannon entropy does not increase by any deterministic computation.

We show the entropy changes in the aforementioned three examples.  We assume that the probability distribution of the input is uniform.
\begin{description}
\item[NOT] $S(M) = S(M') = \ln 2$, and thus $\Delta S(M) = 0$.
\item[ERASE] $S(M) = \ln 2$, $S(M') = 0$, and thus $\Delta S(M) = -\ln 2 < 0$.
\item[AND] $S(M) = 2 \ln 2$, $S(M') = -(3/4) \ln (3/4) - (1/4) \ln (1/4)$, and thus $\Delta S(M) = - (3/4) \ln 3 < 0$.
\end{description}

The equality in the last line of (\ref{S_monotone_proof}) is achieved, if and only if $p(m') = p(m)$ holds for any $(m,m')$ satisfying $m' = \hat C (m)$ and $p(m) \neq 0$.
This is equivalent to the following:  For any $m' \in M'$ with $p(m') \neq 0$, there exists a unique $m \in \hat C^{-1}(m')$ with $p(m) \neq 0$.  This means that $\hat C$ is injection, when the domain of $\hat C$ is restricted to the set of $m \in M$ with  $p(m) \neq 0$.  Therefore, we obtain the following theorem (in a slightly rough expression):\\
\\
\textbf{Theorem.} {\it A deterministic computational process is logically reversible, if and only if the Shannon entropy of the computational process does not change.  
}

\

We are now in a position to discuss why  logical  and thermodynamic reversibilities are fundamentally distinct.
 In fact, logical reversibility is the reversibility of only computational states (i.e., the degrees of freedom (i) in Sec.~\ref{Sec_Introduction}), and thus  characterized by the Shannon-entropy change of computational states, $\Delta S(M)$.
On the other hand, thermodynamic reversibility is the reversibility of the entire system including the heat bath (i.e., the degrees of freedom (i)+(ii)+(iii) in Sec.~\ref{Sec_Introduction}), and thus characterized  by the total entropy production $\Sigma$.  
This observation is summarized in Table \ref{Table_reversibilities}.
We will further develop this observation  in the subsequent sections, especially in the context of the Landauer principle.

\begin{table}
\caption{Characterization of thermodynamic and logical reversibilities.}
\begin{center}
\begin{tabular}{| l || l | l | }\hline
{} & Reversible & Irreversible  \\ \hline \hline
Thermodynamically & $\Sigma = 0$ & $\Sigma > 0$ \\ \hline
Logically & $\Delta S(M) = 0$ & $\Delta S(M) < 0$ \\ \hline
\end{tabular}
\end{center}
\label{Table_reversibilities}
\end{table}

\

We note that any logically irreversible process can be embedded in another logically reversible process by extending the space of computational states~\cite{Bennett}.
For example, if we prepare  $M \times M'$ as an extended set of computational states, we can construct an extended map $\hat C'$ by
\begin{equation}
\hat C' \ :  \  (m,0) \in M \times M'  \ \mapsto \  (m, \hat C(m)) \in M \times M',
\end{equation}
where $0 \in M'$ is the standard state of $M'$.
Strictly speaking, $\hat C'$ should be interpreted as a map from $M \times \{ 0 \}$ to $M \times M'$.
This extended map $\hat C'$ reproduces the original map $\hat C$, if we only look at $M$ of the input and $M'$ of the output.  
A crucial feature of $\hat C'$ is that the input $m \in M$ is kept in $M$ of the output of $\hat C'$.  Therefore, the extended map $\hat C'$ is logically reversible, even when the original map $\hat C$ is logically irreversible.
Such a construction of a logically reversible extension of a logically irreversible map has experimentally be demonstrated in Ref.~\cite{Lopez-Suarez2016} in the context of thermodynamics of computation.

\section{Landauer principle}
\label{Sec_Landauer}

We now discuss thermodynamics of computation.  Before  a general argument, in this section we focus on  a paradigmatic model: the conventional setup of the Landauer principle for information erasure~\cite{Landauer,Shizume1995,Piechocinska2000}.

The information erasure is nothing but the ERASE gate discussed in Sec.~\ref{Sec_Reversibility_computation}: 
The initial information of ``$0$'' or ``$1$'' is erased, so that the final computational state is always in ``$0$'' that is called the standard state.
This is a logically irreversible process as discussed before.

If the initial distribution of the input is uniform (i.e., $p(m=0)=p(m=1)=1/2$),  the Shannon entropy of the input is  $S(M) = \ln 2$, while that of the output is $S(M') = 0$.  The change in the computational entropy is  given by $\Delta S(M) = -\ln 2$, as already shown in Sec.~\ref{Sec_Reversibility_computation}.
This is the erasure of one bit of information.

\begin{figure}[htbp]
 \begin{center}
 \includegraphics[width=120mm]{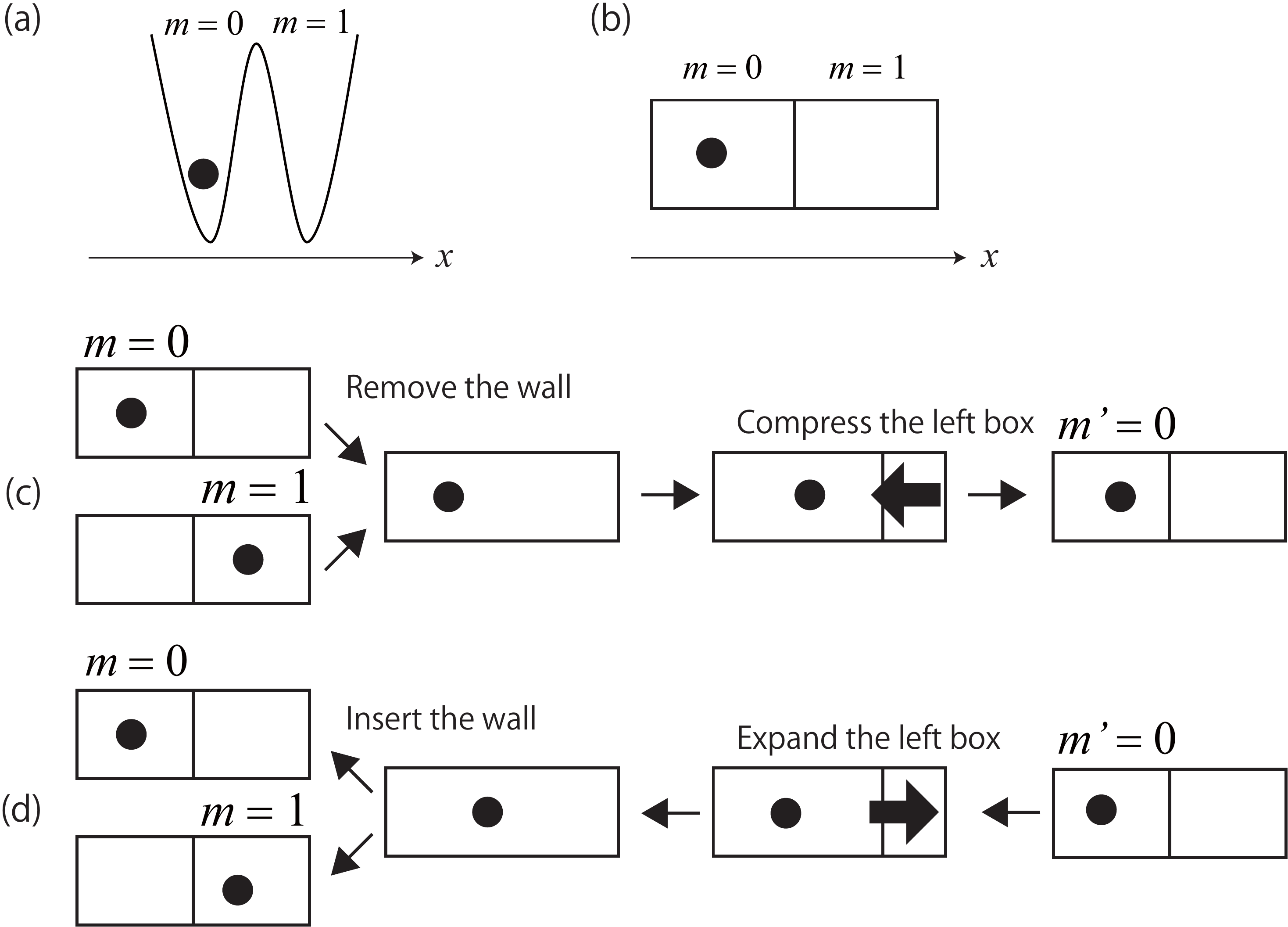}
 \end{center}
 \caption{Schematics of a symmetric memory. (a) The double-well potential model.  The barrier at the center is assumed to be sufficiently high.  (b) The two-box model.   The left (right) box corresponds to the left (right) well of the double-well potential model.
 (c) Schematic of the information-erasure protocol with the two-box model.   The initial computational state is $m=0$ or $m=1$ with probability $1/2$. The wall is instantaneously removed, which does not change the probability distribution of the particle.  Then, the box is compressed  to the left, and the final position of the wall is at the center.  As a consequence, the final computational state  is  the standard state $m'=0$ with unit probability.  
 If the compression process is infinitely slow, this protocol is quasi-static and thermodynamically reversible, where the heat emission is given by $-Q = T \ln 2$.
 (d) The time-reversal of the quasi-static erasure protocol.  The initial computational state is $m'=0$, which is the final computational state of the erasure.
The left box is first expanded infinitely slowly, and then the wall is inserted instantaneously.  The final distribution  is $m=0$ or $m=1$ with probability $1/2$, which is the initial distribution of the erasure.
In this process, the heat absorption is given by $Q = T \ln 2$, which is equal and opposite to that in the erasure process.}   
 \label{Landauer_memory}
\end{figure}

To physically implement information erasure, we consider a physical device that stores one bit of information, which is called a memory.  
Suppose that the memory is in contact with a single heat bath at temperature $T$ ($=\beta^{-1}$).
In the conventional setup, the memory is modeled by a particle in a symmetric double-well potential (see Fig.~\ref{Landauer_memory} (a)), which has already been discussed in Sec.~\ref{Sec_Introduction} (Fig.~\ref{Intro_fig}).
The memory stores ``$0$'' (``$1$''), if the particle is in the left (right) well.
The particle moves stochastically under the effect of a heat bath and can be described by, for example, an overdamped Langevin equation.
Let $x$ be the position of the particle, which is the physical degrees of freedom of this memory.
We assume that the barrier between the wells is sufficiently high compared with the thermal energy $T$ that thermal tunneling between the wells is negligible.

As a simpler model of the memory, the symmetric double-well potential can be replaced by two boxes with an equal volume (Fig.~\ref{Landauer_memory} (b)), where the barrier of the double-well potential corresponds to the wall separating the boxes.
In this two-box model, the memory stores ``$0$'' (``$1$''), if the particle is in the left (right) box.

In any setup (either the double-well model or the two-box model), the entire phase space, which we denote as $X$, represents the position of the particle. 
$X$ is divided into two regions that represent computational states ``$0$'' and ``$1$''.

The information-erasure process with the two-box model is represented in Fig.~\ref{Landauer_memory} (c).
We suppose that the memory is in local equilibrium in the individual boxes in the initial and final distributions.
Since the two boxes have the same volume, the change in the Shannon entropy of the entire phase space by the information erasure is the same as that of the computational states: 
\begin{equation}
\Delta S (X) = \Delta S(M) = -\ln 2.
\end{equation}
From the second law (\ref{second_law2}) with $\Delta S(X) = - \ln 2$, we have
\begin{equation}
-Q \geq T \ln 2,
\label{Landauer1}
\end{equation}
which implies that the heat emission $-Q$ from the memory is bounded from below by $T\ln 2$.
This bound on heat emission is referred to as the Landauer bound, and inequality~(\ref{Landauer1}) is called the Landauer principle.
Experimental verifications of the Landauer principle with a symmetric memory have been performed in, for example, Refs.~\cite{Berut2012,Berut2013,Jun2014,Hong2016,Gavrilov2017}.

Let $W$ be the work performed on the memory during the erasure. Since the internal energy does not change during the erasure, $W  = - Q$ holds from the first law of thermodynamics.  Therefore, the Landauer principle~(\ref{Landauer1}) can be rewritten as
\begin{equation}
W \geq T \ln 2,
\label{Landauer_work}
\end{equation}
which gives  the fundamental lower bound of the work required for the information erasure in a symmetric memory.

The equality in (\ref{Landauer1}) is achieved in the quasi-static limit, where the compression process in Fig.~\ref{Landauer_memory} (c)  is  infinitely slow. 
In fact, such a quasi-static protocol is a special case of the reversible protocol in Fig.~\ref{Reversible_protocol}, and therefore the equality in (\ref{Landauer1}) is achieved from the general argument in Sec.~\ref{Sec_Reversibility_stochastic}.
In this case, the entropy production defined in (\ref{entropy_production2}) is zero: $\Sigma = 0$.
We note that we can also directly compute that $-Q =T\ln 2$ in the infinitely-slow limit, by using the equation of states of the single-particle gas, which will be discussed in Sec.~\ref{Sec_Thermodynamics_computation} for more general situations (the present situation is $t=1/2$ there).

Therefore,  information erasure is thermodynamically reversible in the quasi-static limit.
We note that the probability distribution is unchanged by the removal of the wall, which guarantees that the process is quasi-static. 
On the other hand, if information erasure is not quasi-static, the entropy production is positive: $\Sigma := \Delta S(X) - \beta Q > 0$.  In this case,  information erasure is thermodynamically irreversible.

To be more explicit, we show in Fig.~\ref{Landauer_memory} (d) the time-reversal of the quasi-static information-erasure protocol with the two-box model, which indeed restores the probability distribution to the initial one, leading to the Shannon-entropy change $\Delta S(X)_{\rm reverse} = \ln 2$.
In this time reversal process, the heat of $Q_{\rm reverse} = T\ln 2$ is absorbed from the heat bath during the expansion process, which has the inverse sign of the erasure process. 
We thus confirm that $\Sigma_{\rm reverse} := \Delta S(X)_{\rm reverse} - \beta Q_{\rm reverse} = 0$ in the time-reversal.

In short, logically irreversible information erasure can be performed in a thermodynamically reversible manner.
Of course,  this is totally consistent, given the different definitions of the two reversibilities.
In fact, as also discussed in Sec.~\ref{Sec_Introduction}, logical reversibility cares only about the reversibility of the computational states, while thermodynamic reversibility is characterized by reversibility in the entire universe that consists of the memory and the heat bath.

From the entropic point of view,  logical reversibility implies $\Delta S(M) = 0$, while  thermodynamic reversibility implies $\Sigma := \Delta S(X) - \beta Q = 0$.  
These are definitely different, even when $\Delta S(M) = \Delta S(X)$ as in the present case.

In Table~\ref{Table_reversibilities_conventional}, we summarize the relationship between thermodynamic and logical reversibilities in the standard setup of information erasure.

\begin{table}
\caption{Summary of thermodynamic and logical reversibilities in the conventional setup of information erasure.}
\begin{center}
\begin{tabular}{| l || l | l | }\hline
{} & Quasi-static & Not quasi-static  \\ \hline \hline
Thermodynamically & reversible & irreversible \\ \hline
Logically & irreversible & irreversible \\ \hline
Heat emission $-Q$ & $ = T \ln 2 $ & $>  T \ln 2 $ \\ \hline
Entropy production $\Sigma$ & $= 0$ & $>0$ \\ \hline
\end{tabular}
\end{center}
\label{Table_reversibilities_conventional}
\end{table}

\section{Thermodynamics of computation}
\label{Sec_Thermodynamics_computation}

In this section, we discuss a general framework of  stochastic thermodynamics of computation.
First, we remark that a physical state and a computational state are distinct concepts.
In the standard setup of the Landauer principle in Sec.~\ref{Sec_Landauer}, the physical state is the position of the particle, and thus is a continuous variable, while the computational state is ``left'' or ``right'' of the double well (representing ``$0$'' and ``$1$''), and thus is a binary varibable.
In realistic situations of computation, a single computational state contains a huge number of microscopic physical states.
This can be regarded as a coarse-graining of the physical phase space.

In general, we divide the physical phase space (i.e., the set of physical states) into several non-overlapping regions, where  each region represents  a computational state. 
Let $X$ be the set of physical states, and $M$ be the set of computational states, as in the previous sections.
We consider a subset of $X$, written as $X_m$ with index $m \in M$, which is the set of physical states that represent a computational state $m$. 
$X$ is then divided as $X = \cup_m X_m$, where $X_m  \cap X_{m'} = \phi$ for all $m\neq m'$ with the empty set $\phi$.

We consider a probability distribution on the physical phase space.
Let $P(x)$ be the probability of physical state $x \in X$, and $p(m)$ be that of computational state $m \in M$.  Since all of $x \in X_m$ represent a single computational state $m$, we have
\begin{equation}
p(m) = \sum_{x \in X_m} P(x).
\end{equation}
We then define the conditional probability of $x$ under the condition that the computational state is $m$ (i.e., $x \in X_m$):
\begin{equation}
P(x|m) = \left\{
\begin{array}{l}
P(x)/p(m) \ \ ({\rm if} \ \ x \in X_m), \\
0 \ \ \ ({\rm otherwise}).
\end{array}
\right.
\end{equation}

We next consider the Shannon entropy associated with this probability distribution.
The Shannon entropy of the physical states is given by
\begin{equation}
S(X) := -\sum_x P(x) \ln P(x),
\end{equation}
and the Shannon entropy of the computational states is given by
\begin{equation}
S(M) := - \sum_{m} p(m) \ln p(m).
\end{equation}
We also consider the conditional entropy of $X$ under the condition that the computational state is $m$: 
\begin{equation}
S(X|m) := - \sum_{x \in X_m} P(x|m) \ln P(x|m),
\end{equation}
which represents fluctuations of physical states inside a single computational state.

A crucial property of the non-overlapping devision of the phase space is the  corresponding  decomposition of total (physical) entropy, which is represented as
\begin{equation}
S(X) = S(M) + S(X|M),
\label{decomposition}
\end{equation}
where 
\begin{equation}
S(X|M) := \sum_m p(m) S(X|m) = - \sum_{m \in M} \sum_{x \in X_m} p(m)P(x|m) \ln P(x|m).
\end{equation}
This decomposition is a general property of probability theory~\cite{Cover-Thomas}, and its proof is given by
\begin{equation}
\begin{split}
S(X) &= - \sum_{x \in X} P(x) \ln P(x) \\
&= -\sum_{m \in M} \sum_{x \in X_m} p(m)P(x|m) \ln [p(m) P(x|m)] \\
&= - \sum_{m \in M} p(m) \ln p(m) - \sum_{m \in M} \sum_{x \in X_m} p(m)P(x|m) \ln P(x|m) \\
&=   S(M) + S(X|M),
\end{split}
\end{equation}
where we used $\sum_{x \in X_m} P(x|m) = 1$ to obtain the third line.
We note that $P(x|m) \ln P(x|m) $ does not diverge for all $x$ and $m$.

Here, $S(X)$ represents the entire fluctuation in the physical phase space, which is related to the heat through the second law~(\ref{second_law2}). 
On the other hand, $S(M)$ is the entropy  of computational states, which is related to logical reversibility as discussed in Sec.~\ref{Sec_Reversibility_computation}.
$S(X|M)$ represents the average of fluctuations inside the individual computational states.
We refer to $S(X)$ as the physical entropy, $S(M)$ as the computational entropy, and $S(X|M)$ as the internal entropy.

We next consider dynamics on $X$ that realizes a computation $\hat C$.
The dynamics can be stochastic on the entire phase space $X$ but should be deterministic on the computational space $M$ in order to realize a deterministic computation.
Such a situation is realistic in practical computations, because physical states thermally fluctuate inside individual computational states, even when the output of computation is deterministic.

We consider the change in the entropy during computation.
Let $m$ and $m'$ be the initial and final computational states that are related deterministically as  $m' = \hat C (m)$, and  $x$ and $x'$ be the initial and final physical states that are related stochastically.
We use notations $X$ and $X'$ ($M$ and $M'$) to refer to the probability variables of the initial and final physical (computational) states, respectively.
The change in the Shannon entropies are then denoted as $\Delta S(X) := S(X') - S(X)$, $\Delta S(M) := S(M') - S(M)$, and $\Delta S(X|M) := S(X' | M' ) - S(X|M)$.

The second law (\ref{second_law2}) is represented by the total entropy as
\begin{equation}
\Delta S(X) \geq \beta Q,
\label{second_law3}
\end{equation}
which is equivalent to, via decomposition (\ref{decomposition}),
\begin{equation}
\Delta S(M) + \Delta S(X|M) \geq \beta Q. 
\label{second_law4}
\end{equation}
Correspondingly, the entropy production (\ref{entropy_production2}) is decomposed as
\begin{equation}
\Sigma = \Delta S(M) + \Delta S(X|M)  - \beta Q.
\end{equation}

In the present setup, the nonequilibrium free energy is the same as Eq.~(\ref{noneq_free}):
\begin{equation}
F(X) := E(X) - TS(X) = E(X) - TS(M) - TS(X|M),
\label{noneq_free_memory} 
\end{equation}
where $E(X)$ is the average energy of the memory.  Then, the fundamental lower bound of the work $W$ required for the computation is given by
\begin{equation}
W \geq \Delta F(X).
\label{noneq_work_memory}
\end{equation}

We consider the  local canonical distribution inside a computational state $m$, which is given by
\begin{equation}
P_{\rm can}(x|m) := 
\left\{
\begin{array}{l}
e^{\beta (F_{\rm eq}(m) - E(x|m))} \ \ ({\rm if} \ \ x \in X_m), \\
0 \ \ \ ({\rm otherwise}),
\end{array}
\right.
\end{equation}
where $E(x|m)$ is the Hamiltonian for a given $m$, and 
\begin{equation}
F_{\rm eq}(m) := - T \ln \sum_{x \in X_m} e^{-\beta E(x|m)}
\end{equation}
is the local equilibrium free energy under the condition of $m$.
If the memory is in local equilibrium inside individual computational states, the nonequilibrium free energy (\ref{noneq_free_memory}) reduces to~\cite{Sagawa-Ueda2009}
\begin{equation}
F(X)  = F_{\rm eq} - TS(M),
\end{equation}
where $F_{\rm eq} := \sum_{m} p(m) F_{\rm eq}(m)$.  If the initial and final distributions are local canonical, inequality (\ref{noneq_work_memory}) reduces to
\begin{equation}
W \geq \Delta F_{\rm eq} - T\Delta S(M).
\label{noneq_work_memory2}
\end{equation}

In the rest of this section, we assume that the initial and final distributions are local canonical.  In fact, this is a reasonable assumption, given that the time scale of global thermalization is much longer than that of local thermalization, because of the potential wall between the computational states.

\

$\Diamond\Diamond\Diamond$

\

We now consider the role of the symmetry of the memory.
We first consider the case that the memory is symmetric as in Sec.~\ref{Sec_Landauer}.
In such a case, the local free energies of the two computational states are the same: $F_{\rm eq} (0) = F_{\rm eq}(1)$, and therefore $\Delta F_{\rm eq} = 0$ for any computation.
Therefore, inequality (\ref{noneq_work_memory2}) reduces to
\begin{equation}
W \geq - T\Delta S(M),
\label{noneq_work_memory3}
\end{equation}
which is a general expression of the Landauer principle.
In fact, the original Landauer principle (\ref{Landauer_work}) is a special case of inequality~(\ref{noneq_work_memory3}) with $\Delta S(M) = - \ln 2$.
Inequality~(\ref{noneq_work_memory3}) has been directly verified in a recent experiment~\cite{Gavrilov2017}.

In terms of the internal entropy, the symmetry implies $S(X|0) = S(X|1)$ in local equilibrium, and therefore $\Delta S(X|M) = 0$.   Then, inequality~(\ref{second_law4}) reduces to
\begin{equation}
\Delta S(M)  \geq \beta Q,
\label{Landauer2}
\end{equation}
or equivalently,
\begin{equation}
- Q \geq  - T \Delta S(M). 
\label{Landauer3}
\end{equation}

\begin{figure}[htbp]
 \begin{center}
 \includegraphics[width=120mm]{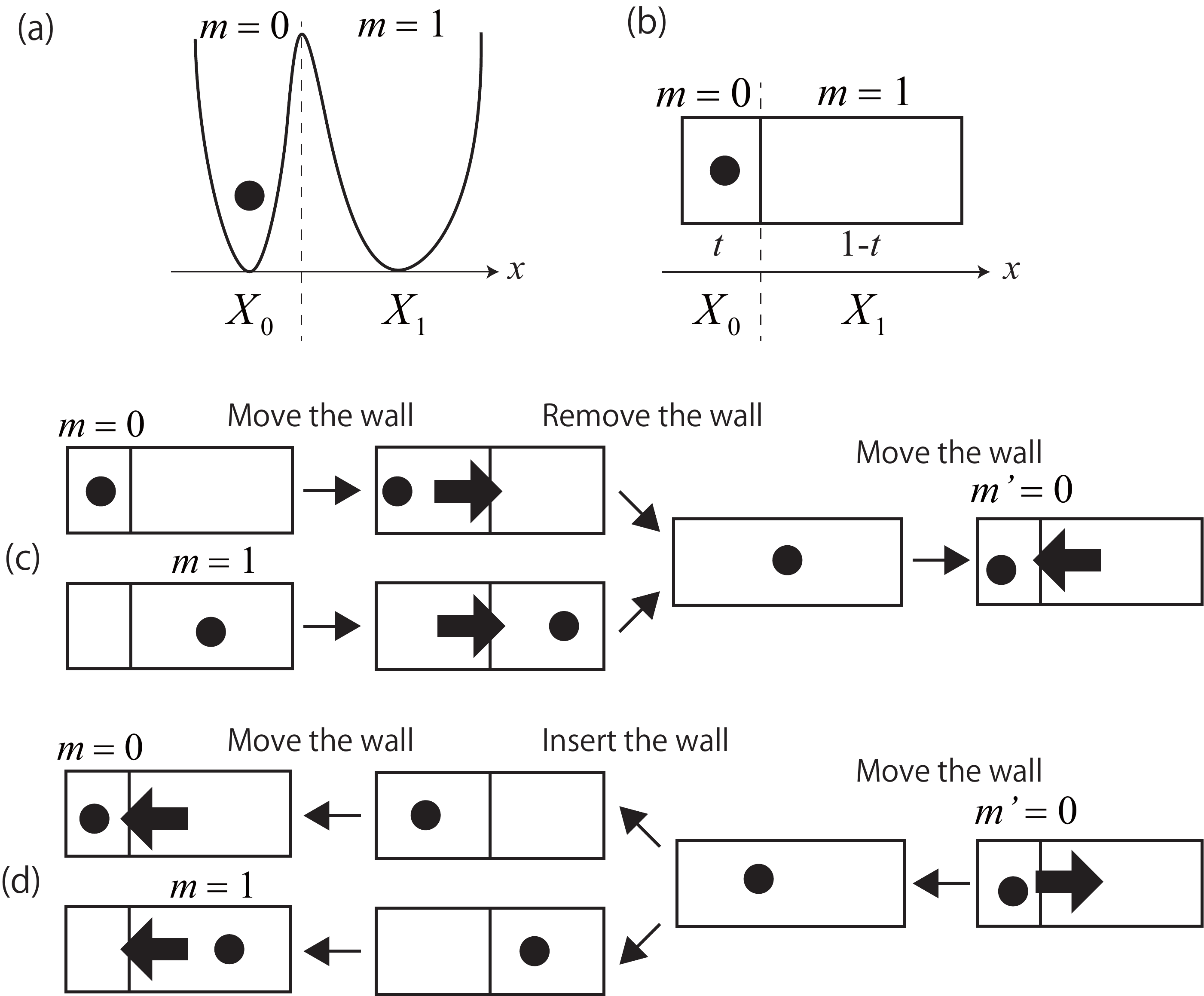}
 \end{center}
 \caption{Schematics of an asymmetric memory~\cite{Sagawa-Ueda2009}. (a) An asymmetric double-well model.  The phase-space volume of $X_0$ and $X_1$ are not equal.  (b) The corresponding two-box model.  The volumes of the left and right boxes are not equal. The volume ratio of the two boxes is given by $t:1-t$ ($0<t<1$).  (c) The optimal information-erasure protocol with the asymmetric two-box model, which achieves the thermodynamic reversible condition $\Sigma = 0$.  The initial computational state is  $m=0$ or $m=1$ with probability $1/2$.  The wall is moved to the center of the box infinitely slowly, and then removed instantaneously.  The box is then compressed to the left infinitely slowly, so that  the final volume ratio of the two boxes is the same as the initial one.  The entire process of this erasure is quasi-static.
(d) The time-reversal of the above quasi-static erasure protocol.  The initial distribution of the time-reversal is the same as the final distribution of the erasure.  The wall is first moved to the right most position infinitely slowly, and then a wall is instantaneously inserted at the center of the box.  The inserted wall is then moved infinitely slowly, such that its final position is the same as its initial position of the erasure.  
In this process, the total heat absorption is given by $Q = T\ln 2 + (T/2) \ln (1-t/t)$, which is equal and opposite to  that in the erasure process. }  
\label{Landauer_memory_asymmetric}
\end{figure}

We next consider the case that the memory is asymmetric, where $\Delta F_{\rm eq} \neq 0$ and  $\Delta S(X|M) \neq 0$ in general.
If $\Delta S(X|M) \neq 0$, the entropy change in the computational states is not directly related to the heat (i.e., inequality (\ref{Landauer2}) does not necessarily hold)~\cite{Barkeshli2005,Turgut2009,Sagawa-Ueda2009}.
In Fig.~\ref{Landauer_memory_asymmetric} (a), we show a simple example of a memory with an asymmetric double-well potential, where the left (right) well represents computational state ``$0$'' (``$1$'').

As is the case for the symmetric memory, we can replace the double-well potential by two boxes (Fig.~\ref{Landauer_memory_asymmetric} (b)).
If the double-well potential is asymmetric, the volumes of the two boxes are not the same.
Let $t$ ($0 < t < 1$) be the ratio of the volume of the left box.  
If the memory is symmetric, $t=1/2$.
For $0<t<1/2$, the local free energies satisfy $F_{\rm eq} (0) > F_{\rm eq} (1)$, and the internal entropies satisfy $S(X|0) < S(X|1)$ in local equilibrium.
We emphasize that the initial probability distribution of $m=0$ and $m=1$ is arbitrary (i.e., not necessarily $p(m=0)=t$), because the memory can store any information.

We consider information erasure with the asymmetric memory.
For simplicity,  we assume that the initial distribution is $p(m=0) = p(m=1) = 1/2$.
The Shannon-entropy change in the computational states by the information erasure is then given by $\Delta S(M) = -\ln 2$.

Figure \ref{Landauer_memory_asymmetric} (c) shows  the optimal information-erasure protocol, which achieves the equality of the second law~(\ref{second_law4})~\cite{Sagawa-Ueda2009}.
A crucial point of this protocol is that the wall is first moved to the center infinitely slowly.
Thanks to this process, the probability distribution of the particle (i.e., $1/2$ for both left and right) does not change by the removal of the wall.
(If we removed the wall without moving it to the center, the probability distribution would spontaneously relax towards the uniform distribution over the box, which makes the process thermodynamically irreversible and the entropy production positive.)
This is in the same spirit as the protocol in Fig.~\ref{Reversible_protocol}.
Then, the box is compressed from to the left infinitely slowly, and the final position of the wall returns to the initial one.
The total entropy production is zero in this process, and thus it is thermodynamically reversible.
To see the thermodynamic reversibility more explicitly, we illustrate the time-reversal of the above protocol in Fig.~\ref{Landauer_memory_asymmetric} (d).

We can also directly compute the heat emission for the protocol in Fig.~\ref{Landauer_memory_asymmetric} (c).
We assume the equation of states of the single-particle gas, i.e., $PV = T$, where $P$ is the pressure and $V$ is the volume of the box (and remind that $k_{\rm B}= 1$).
The heat emission is then given by $- Q = W = - \int PdV$.  
We note that the work is not needed for the removal of the wall.
Then, the heat emission during the entire process is given by
\begin{equation}
-Q = T\ln 2 + \frac{T}{2} \ln \frac{1-t}{t}.
\label{Q_asymmetric}
\end{equation}
On the other hand, we have $S(X | m= 0) = S(X' | m' = 0)$ and
\begin{equation}
S(X' | m' = 0) - S(X|m=1) = \ln \frac{t}{1-t},
\end{equation}
and therefore,
\begin{equation}
\Delta S(X|M) := S(X'|m'=0) - \frac{1}{2} \left( S(X|m=0) + S(X|m=1) \right) = \frac{1}{2} \ln \frac{t}{1-t}.
\end{equation}
Combining this with $\Delta S(M) = -\ln 2$, we obtain
\begin{equation}
\Delta S(M) + \Delta S(X|M) - \beta Q =  - \ln 2 + \frac{1}{2} \ln \frac{t}{1-t}  +\left( \ln 2 + \frac{1}{2} \ln \frac{1-t}{t} \right) = 0,
\end{equation}
which achieves the equality in  (\ref{second_law4}).

If $t=1/2$, Eq.~(\ref{Q_asymmetric}) reproduces that $-Q = T\ln 2$.
On the other hand,  if $t \neq 1/2$,  we have $-Q \neq T\ln 2$.
In particular, if $t> 1/2$, we have $-Q < T \ln 2$, which is  below the Landauer bound (\ref{Landauer3}).
Of course, this does not contradict the second law.
In such a case, the decrease in the computational entropy $\Delta S(M)$ is compensated for by the increase in the internal entropy $\Delta S(X|M)$.
Information erasure with such an asymmetric memory has experimentally been demonstrated in Ref.~\cite{Gavrilov2016a}.

In summary,  heat emission is connected to the change in the total physical entropy of the memory (i.e., (i)+(ii) in Sec.~\ref{Sec_Introduction}), which is decomposed into the computational and internal entropies as in Eq.~(\ref{decomposition}).
If the change in the internal entropy is not zero, the computational entropy is not directly related to heat emission.
This is the reason why the information erasure below the Landauer bound (\ref{Landauer3}) is possible with an asymmetric memory, while the general bound (\ref{second_law4}) is always true.

\section{Work extraction and reversibility with feedback control}
\label{Sec_Work}

We next consider work extraction from heat engines through feedback control by Maxwell's demon. 
As we will discuss below, the mutual information is the source of work extraction by the demon, and therefore we refer to such a heat engine as an information engine.

 In this section, we do not explicitly formulate the memory of the demon itself.  Instead, we only regard the demon as an external agent that affects the probability distribution of the engine through the measurement.  The full analysis of the total system of engine and demon is postponed to the next section.

\begin{figure}[htbp]
 \begin{center}
 \includegraphics[width=100mm]{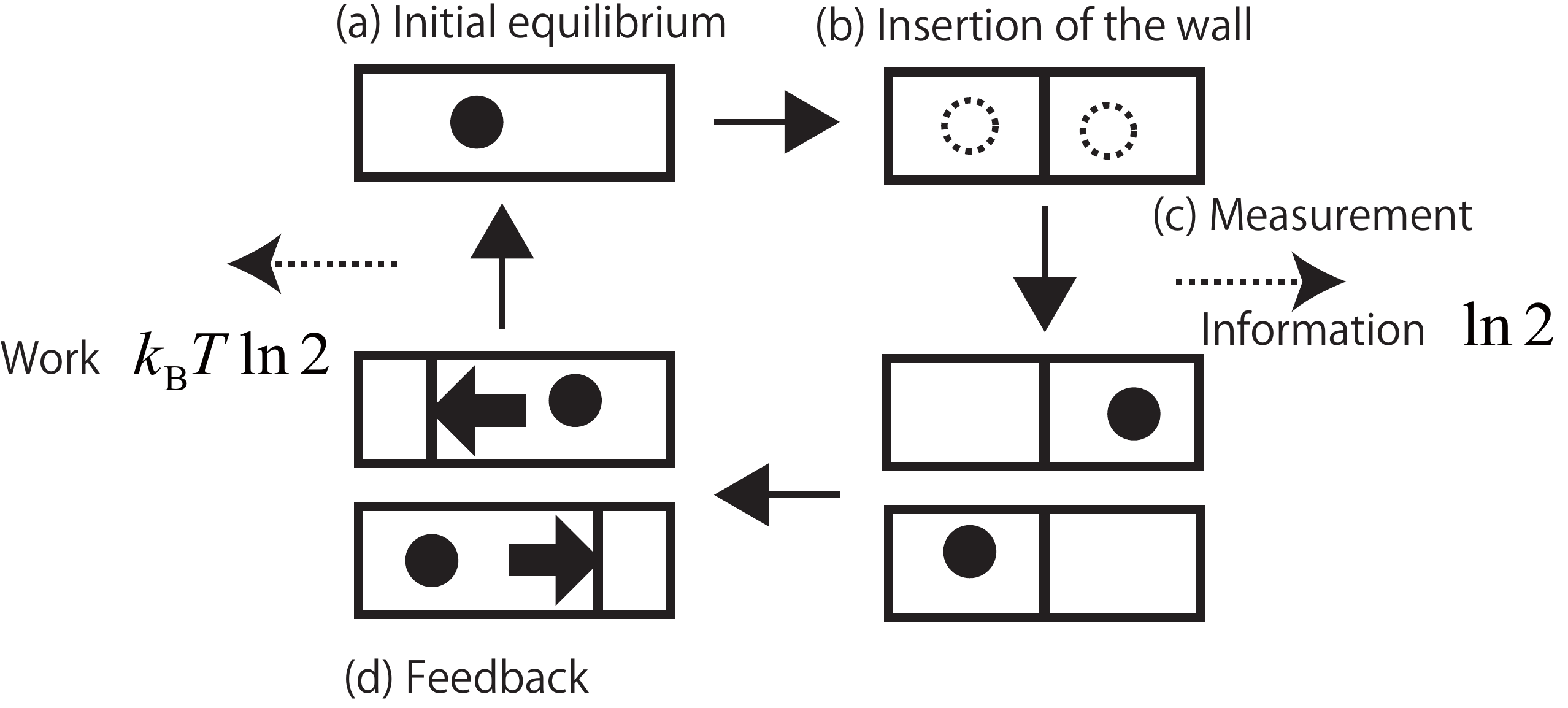}
 \end{center}
 \caption{Schematic of the Szilard engine.  (a) The initial equilibrium distribution.  (b)  A wall  is inserted at the center of the box. At this stage, we do not know in which side the particle is.  (c) The demon measures the position of the particle (i.e., left or right).  (d)  If the particle of the engine is found in the left (right) box, the demon infinitely slowly expands the box to the right (left) so that the final distribution returns to the initial one.  The work of $W_{\rm ext} = T \ln 2$ is extracted from this expansion.  Since the direction of the expansion depends on the measurement outcome (left or right), this process is regarded as feedback control by the demon.} 
 \label{Szilard}
\end{figure}

We first consider the Szilard engine, which is a simple model of an information engine (see Fig.~\ref{Szilard}).
The Szilard engine consists of  a Brownian particle (or a molecule) in a box that is attached to a single heat bath at temperature $T = \beta^{-1}$.
During the process depicted in Fig.~\ref{Szilard}, the demon obtains one bit ($= \ln 2$) of information corresponding to left or right, and performs feedback control.
Then, the work of $W_{\rm ext} = T \ln 2 > 0$ is extracted from the engine, and the same amount of heat $Q = T \ln 2$ is absorbed from the bath.  The amount of the work is calculated by the same manner as in Sec.~\ref{Sec_Thermodynamics_computation}. 
Since the dynamics of the engine is cyclic, this positive work extraction  apparently violates the second law.
However, if we take into account the memory of the demon, then the total system is not cyclic, and therefore this is not a violation of the second law. 
As will be discussed in Sec.~\ref{Sec_Entropy}, we can understand more quantitatively the consistency between the demon and the second law by taking into account the mutual information between the engine and the demon.

We now consider a general upper bound of the extractable work with feedback control.
We assume that the engine is in contact with a single heat bath at temperature $T = \beta^{-1}$.
Let $P(x)$ be the probability distribution of the engine immediately before the measurement by the demon. 
We note that we use notation $x$ (and $X$) to describe the engine, instead of the memory of the demon; we use notation $y$ for the measurement outcome obtained by the demon. 

 We suppose that the error of the measurement is described by the conditional probability $P(y|x)$, which is the probability  of obtaining outcome $y$ under the condition that the true state of the engine is $x$.
If the measurement is error-free, we have $P(y|x) = \delta_{xy}$ with $\delta_{xy}$ being the Kronecker's delta (or the delta function if the state variable is continuous).
The  joint probability of $x$ and $y$ is given by $P(x,y) = P(y|x)P(x)$, and the unconditional probability of $y$ is given by $P(y) = \sum_x P(x,y)$.
From the Bayes rule, the conditional probability of $x$ under the condition of outcome $y$ is given by
\begin{equation}
P(x|y) = \frac{P(x,y)}{P(y)}.
\end{equation}
Correspondingly, the conditional entropy of $X$ under the condition of a particular $y$ is given by
\begin{equation}
S(X|y) := - \sum_x P(x|y) \ln P(x|y).
\end{equation}
Its ensemble average over all $y$ is 
\begin{equation}
S(X|Y) := \sum_y P(y) S(X|y) = - \sum_{xy} P(x,y) \ln P(x|y) = S(XY) - S(Y),
\end{equation}
where $S(XY) := - \sum_{xy} P(x,y)\ln P(x,y)$ is the Shannon information of the joint distribution.

After the measurement, the protocol to control the engine depends on $y$, which is the characteristic of feedback control.
By noting that  the initial distribution of the engine is given by $P(x|y)$ under the condition of outcome $y$,  the second law of stochastic thermodynamics (\ref{second_law2}) can apply to the conditional distribution:
\begin{equation}
S(X' | y) - S(X|y) \geq \beta Q_y,
\label{second_law_feedback0}
\end{equation}
where $Q_y$ is the heat absorption with $y$, and $S(X'|y)$ is the conditional entropy in the final distribution of the engine.
By taking the ensemble average over all $y$, we have
\begin{equation}
S(X'|Y) - S(X|Y) \geq \beta Q,
\label{second_law_feedback1}
\end{equation}
where $Q:=\sum_y P(y) Q_y$.

Before proceeding further, we here discuss  mutual information, which quantifies a correlation between two probability variables.
The mutual information between $X$ and $Y$ is defined as
\begin{equation}
I(X:Y) := S(X) + S(Y) - S(XY) = \sum_{x,y} P(x,y) \ln \frac{P(x,y)}{P(x)P(y)}.
\end{equation}
It immediately follows that
\begin{equation}
I(X:Y) =  S(X) - S(X|Y) = S(Y) - S(Y|X).
\label{mutual_conditional}
\end{equation}
The mutual information satisfies the following inequalities:
\begin{equation}
0 \leq I(X:Y) \leq \min \{ S(X), S(Y) \},
\end{equation}
where $I(X:Y) = 0$ holds if and only if the two systems are not correlated (i.e., $P(x,y) =P(x)P(y)$).

Going back to the second law (\ref{second_law_feedback1}), it is rewritten as, by using Eq.~(\ref{mutual_conditional}),
\begin{equation}
\Delta S(X) -  \beta Q \geq - \Delta I,
\label{second_law_feedback2}
\end{equation}
where $\Delta S(X) := S(X') - S(X)$ and $\Delta I := I(X':Y) - I(X:Y)$.
We note that if the feedback control works, $I(X':Y) < I(X:Y)$ should hold (and thus $\Delta I < 0$).  
 In fact, in the case of the Szilard engine, $I(X:Y) = \ln 2$ and $I(X':Y)= 0$ hold, because there is no remaining correlation after the entire process.
In general, since the correlation is also decreased by dissipation to the environment, $-\Delta I$  gives an upper bound of the information that is utilized by feedback control.
By noting that $\Delta S(X) -  \beta Q$ is nonnegative  in the absence of  feedback control, inequality~(\ref{second_law_feedback2}) implies that we can reduce the entropy of the system by using feedback control, where the mutual information is the resource of the entropy reduction.

We consider  the nonequilibrium free energy of $X$, defined in the same manner as Eq.~(\ref{noneq_free}):
\begin{equation}
F(X) :=  E(X) -T  S(X),
\end{equation}
where $E(X)$ is the average energy of the engine.  We then rewrite inequality (\ref{second_law_feedback2}) as
\begin{equation}
W \geq \Delta F(X) + T \Delta I.
\label{second_law_feedback3}
\end{equation}
By defining the extracted work $W_{\rm ext} := -W$, we have
\begin{equation}
W_{\rm ext} \leq - \Delta F(X) - T \Delta I.
\end{equation}
The right-hand side above can be further bounded as
\begin{equation} 
W_{\rm ext} \leq - \Delta F(X) + TI(X:Y),
\label{second_law_feedback4}
\end{equation}
where we used $I(X':Y) \geq 0$.  Inequality~(\ref{second_law_feedback4}) implies that additional work can be extracted up to the mutual information obtained by the measurement.

We consider a special case that the initial distribution of the engine  is canonical  and the final distribution is also canonical under the condition of $y$.  More precisely, the final Hamiltonian can depend on $y$, which we denote by $E(x'|y)$, and the final distribution is given by $P(x'|y) = e^{\beta (F_{\rm eq} (X'|y)-E(x'|y))}$, where $F_{\rm eq} (X'|y) := - T \ln \sum_{x'}e^{-\beta E(x'|y)}$ is the final equilibrium free energy with $y$.    Let $F_{\rm eq}(X)$ be the initial equilibrium free energy as usual .  We then have
\begin{equation}
S(X) = \beta ( E(X) - F_{\rm eq}(X)), \ S(X'|y) = \beta (E(X'|y) - F_{\rm eq}(X'|y)),
\label{eq_free_energy_feedback}
\end{equation}
where $E(X'|y) := \sum_{x'} P(x'|y)E(x'|y)$ that gives $E(X') = \sum_y P(y) E(X'|y)$.
We define the change in the equilibrium free energy by
\begin{equation}
\Delta F_{\rm eq} (X) := \sum_y P(y) F_{\rm eq} (X'|y) - F_{\rm eq}(X).
\label{eq_free_energy_change}
\end{equation}
By substituting Eq.~(\ref{eq_free_energy_feedback}) into inequality~(\ref{second_law_feedback3}), we have
\begin{equation}
W \geq  \Delta F_{\rm eq} (X) - TI(X:Y), 
\label{second_law_feedback5}
\end{equation}
or equivalently,
\begin{equation}
W_{\rm ext} \leq   - \Delta F_{\rm eq} (X) + TI(X:Y).
\label{second_law_feedback6}
\end{equation}
We emphasize that inequality (\ref{second_law_feedback5}) or (\ref{second_law_feedback6}) is exactly equivalent to (\ref{second_law_feedback3}) under the assumption that the initial and final distributions are (conditional) canonical, where we did not drop $T I(X':Y)$.   In fact, to obtain inequality (\ref{second_law_feedback5}) or (\ref{second_law_feedback6}) from  (\ref{second_law_feedback3}),    $T I(X':Y)$ is just absorbed into the definition of $\Delta F_{\rm eq} (X)$ in Eq.~(\ref{eq_free_energy_change}).
On the other hand, we dropped $T I(X':Y)$ to obtain (\ref{second_law_feedback4})  from  (\ref{second_law_feedback3}).

In the case of the Szilard engine, we have $W_{\rm ext} = T \ln 2$, $\Delta F_{\rm eq} (X) = 0$, and $I(X:Y) = \ln 2$.  Therefore, the equality in  (\ref{second_law_feedback6}) is achieved in the Szilard engine.  

We note that inequality (\ref{second_law_feedback6}) has been derived in Refs.~\cite{Sagawa-Ueda2008,Sagawa-Ueda2010}.
The role of mutual information in thermodynamics has been experimentally demonstrated in, for example, Refs.~\cite{Koski2014,Camati2016}.

\

$\Diamond\Diamond\Diamond$

\

We  consider thermodynamic reversibility with feedback control~\cite{Jacobs2009,Horowitz2011a}.
We remember that the second law with feedback control is given by inequality~(\ref{second_law_feedback1}), which is the ensemble average of inequality (\ref{second_law_feedback0}).  Here,  inequality (\ref{second_law_feedback0}) is equivalent to the second law (\ref{second_law2}) under the condition of $y$.  Therefore, it is reasonable to adopt the following definition~\cite{Horowitz2011a}:\\
\\
\textbf{Definition (Thermodynamic reversibility with feedback).}
\textit{In the presence of feedback control, thermodynamic reversibility of the engine is achieved if and only if the equality in (\ref{second_law_feedback1}) is achieved, or equivalently, the equality in (\ref{second_law_feedback0}) is achieved for all $y$.}  

\

In the rest of this section, we work on this definition of thermodynamic reversibility.  We note, however, that this definition does not concern the reversibility of the memory of the demon during the measurement process.  In fact, in this section we have just formulated the measurement process as the modification of the probability distribution from $P(x)$ to $P(x|y)$, without explicitly considering dynamics of the memory.  The full treatment of the memory  will be discussed in Sec.~\ref{Sec_Entropy} in detail.

By remembering the argument in Sec.~\ref{Sec_Reversibility_stochastic}, thermodynamic reversibility is achieved by the protocol in Fig.~\ref{Reversible_protocol}, where we now replace the distribution $P(x)$ by the conditional one $P(x|y)$, and also the potential $V(x)$ by the $y$-dependent one $V(x|y)$.  In other words, thermodynamic reversibility with feedback control is achieved if we adjust the potential $V(x|y)$ such that the conditional distribution $P(x|y)$ becomes always the canonical distribution of $V(x|y)$.
In particular, we need to switch the potential immediately after the measurement, because the distribution is suddenly changed from $P(x)$ to $P(x|y)$ by the measurement.
We again remark that this consideration neglects reversibility of the measurement process itself.

We revisit the Szilard engine as a special example.  Since the Szilard engine achieves the equality in  (\ref{second_law_feedback6}) as mentioned above, the Szilard engine is thermodynamically reversible.  We can directly confirm that the Szilard engine is always in the canonical distribution under a particular measurement outcome. 

\begin{figure}[htbp]
 \begin{center}
 \includegraphics[width=120mm]{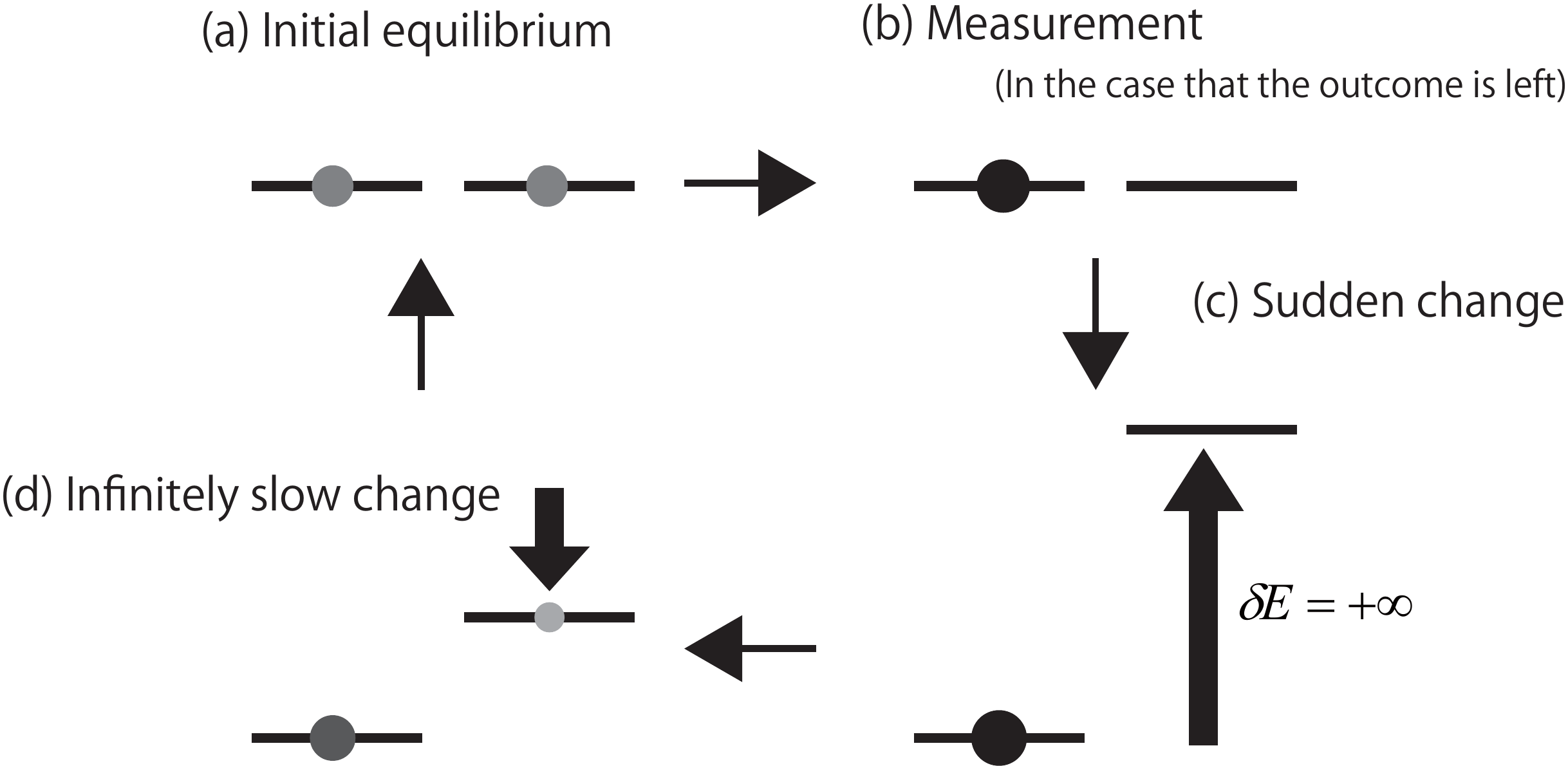}
 \end{center}
 \caption{An analogue of the Szilard engine.  (a) In the initial equilibrium distribution, the particle is in the left or the right site with probability $1/2$.  (b) The demon performs the measurement of the position of the particle, and obtains $\ln 2$ of information. The case that the outcome is ``left'' is shown in this figure.  (c)  If the particle is found in the left (right) site, the demon suddenly changes the energy level of the right (left) site to $+\infty$.  This is analogous to the insertion of the wall of the original Szilard engine.  In this sudden change, we do not need any work.  (d) The demon infintely slowly lowers the energy level of the right (left) site to the original level, from which  $T \ln 2$ of the work is extracted.} 
 \label{Szilard2}
\end{figure}

 To see this point clearer, let us consider a simple analogue of the Szilard engine, illustrated in Fig.~\ref{Szilard2}.  
In this model, the particle is in one of the two sites with the same energy, which is in contact with a single heat bath at temperature $T$ ($= \beta^{-1}$).  
The information gain $\ln 2$ and the work extraction $T \ln 2$ in this model are the same as those in the Szilard engine, implying the thermodynamic reversibility of this model.  It is obvious that this model is always in the conditional canonical distribution during the entire process.

\begin{figure}[htbp]
 \begin{center}
 \includegraphics[width=120mm]{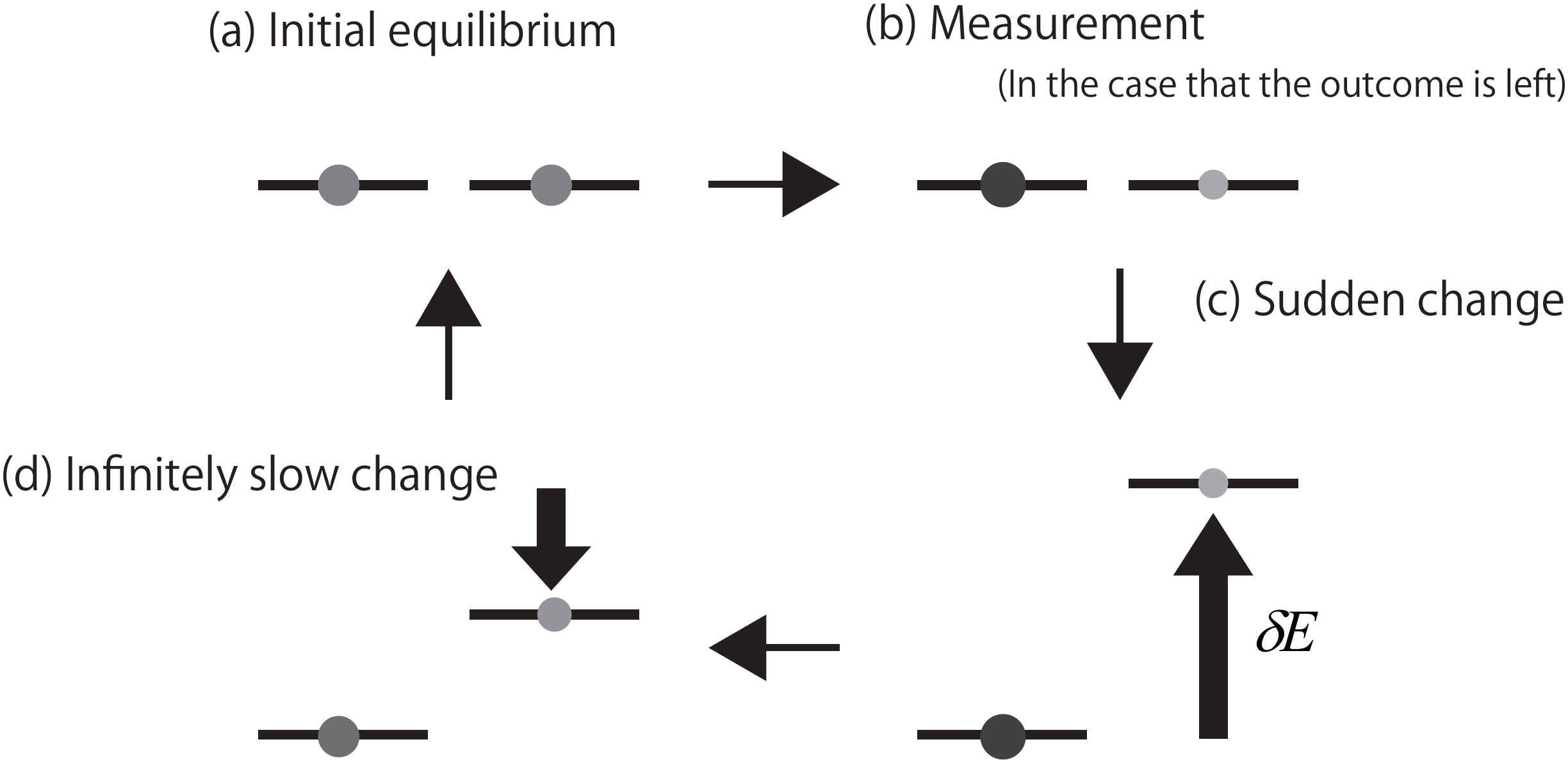}
 \end{center}
 \caption{The Szilard-type engine with measurement error~\cite{Horowitz2011a}. (a)  In the initial equilibrium distribution, the particle is in the left or the right site with probability $1/2$.  (b) The demon performs the measurement of the position of the particle, and obtains the mutual information (\ref{Szilard_mutual}). The case that the outcome is ``left'' is shown in this figure.  (c)  If the particle is found in the left (right) site, the demon suddenly changes the energy level of the right (left) site such that the energy difference is given by $\delta E$.  For this sudden change,  a positive amount of work is performed if $\varepsilon \neq 0$. (d) The demon infiinitely slowly lowers the energy level of the right (left) site to the original level, from which  a positive amount of work is extracted.  } 
\label{Szilard_error}
\end{figure}

We can generalize this model by incorporating a measurement error~\cite{Horowitz2011a}, which is illustrated in Fig.~\ref{Szilard_error}.
We suppose that the error rate of the measurement is given by $\varepsilon$ ($0\leq \varepsilon \leq 1$); the conditional probabilities are given by $P(y|x) = 1-\varepsilon$ ($x = y$) and $P(y|x) =\varepsilon$ ($x \neq y$) with $x,y$ being ``right'' or ``left''.  In this case, the mutual information obtained by this measurement is
\begin{equation}
I(X:Y) = \ln 2 + \varepsilon \ln \varepsilon + (1-\varepsilon) \ln (1-\varepsilon). 
\label{Szilard_mutual}
\end{equation}

 Immediately after the measurement, we have  $P(x|y) = 1-\varepsilon$ ($x = y$) and $P(x|y) =\varepsilon$ ($x \neq y$). To achieve thermodynamic reversibility, we need to make $P(x|y)$ the canonical distribution for all $y$.  Consider the case that  $y=$``left'' as illustrated in Fig.~\ref{Szilard_error}.   (The same argument applies to the case that $y=$``right''.)  The demon switches the energy level of the right site to make the energy difference $\delta E = - T \ln (\varepsilon / (1-\varepsilon))$ so that $P(x|y)$ becomes canonical (Fig.~\ref{Szilard_error} (c)):
\begin{equation}
\frac{e^{-\beta \delta E}}{1+e^{-\beta \delta E}} = \varepsilon, \ \frac{1}{1+e^{-\beta \delta E}} = 1- \varepsilon.
\end{equation}
The work extraction by this switching is given by $- \varepsilon \delta E$ on average, because the particle is pushed up if it is in the right site.

The demon next lowers the energy level of the right site infinitely slowly, and the final distribution is the same as the initial one (Fig.~\ref{Szilard_error} (d)).
The extracted work during this process is given by $T \ln (2/(1+e^{-\beta \delta E}))$, because the extracted work equals the minus of the equilibrium free-energy change in this situation (i.e., the free energy after the sudden switching is $-T\ln (1+e^{-\beta \delta E})$ and that in the final distribution is $-T \ln 2$).

The total work extracted from the entire process is then given by
\begin{equation}
W_{\rm ext} = - \varepsilon \delta E + T \ln \frac{1}{1+e^{-\beta \delta E}} = T \left( \ln 2 + \varepsilon \ln \varepsilon + (1-\varepsilon) \ln (1-\varepsilon) \right).
\end{equation}
We note that $\Delta F_{\rm eq} (X)= 0$ in the entire process.  Therefore, the equality in (\ref{second_law_feedback6}) is achieved, i.e., $W_{\rm ext} = TI(X:Y)$, and thus we confirm that this protocol achieves the thermodynamic reversibility.

This type of the Szilard engine with measurement error has been proposed in Ref.~\cite{Horowitz2011a}, and experimentally demonstrated in Ref.~\cite{Koski2014} by using a single electron box.
Other models that achieve the thermodynamic reversibility with feedback have been discussed in Refs.~\cite{Horowitz2011b,Abreu2011,Horowitz2013,Kwon2017,Sagawa-Ueda2012a}.

\section{Entropy balance in Maxwell's demon}
\label{Sec_Entropy}

In the previous section, we did not explicitly consider the measurement as a physical process, but as the modification of the probability distribution from $P(x)$ to $P(x|y)$.  In particular, we did not consider the entropy production in the memory of the demon itself.

In this section, we explicitly consider stochastic thermodynamics of the entire system of the engine $X$ and the memory of the demon $Y$~\cite{Sagawa-Ueda2012b,Sagawa-Ueda2013}.  Specifically, we focus on the entropy balance during the measurement and the feedback processes by explicitly considering the memory as a physical system.
In this respect, we will reproduce the second law~(\ref{second_law_feedback1}) with feedback control from a slightly different viewpoint from Sec.~\ref{Sec_Work}.  We also discuss the fundamental energy cost required for the measurement process.

As a preliminary, we consider general dynamics of the bipartite system $X$ and $Y$ in the presence of a heat bath at temperature $T = \beta^{-1}$.
The entropy production in the total system is given by
\begin{equation}
\Sigma (XY) := \Delta S(XY) - \beta Q_{XY},
\end{equation}
where $\Delta S(XY)$ is the change in the joint Shannon entropy, and $Q_{XY}$ is the heat absorbed by the total system. 
We can also define the entropy production in the subsystem $X$ by
\begin{equation}
\Sigma (X) := \Delta S(X) -\beta Q_X,
\end{equation}
where $\Delta S(X) $ is the change in the Shannon entropy of $X$, and $Q_X$ is the heat absorbed by $X$ from the heat bath.  In the same manner, we define
\begin{equation}
\Sigma (Y) := \Delta S(Y) -\beta Q_Y.
\end{equation}

In many physical situations (e.g., a bipartite Markov jump process and a Langevin system with two variables driven by independent noise), we can suppose  that the heat is additive:
\begin{equation}
Q_{XY} = Q_X + Q_Y.
\label{heat_additive}
\end{equation}
On the other hand, the Shannon entropy  is generally not additive, and the mutual information appears:  
\begin{equation}
\Delta S (XY) = \Delta S (X) + \Delta S (Y) - \Delta I(X:Y).
\label{Shannon_additive}
\end{equation}
By using Eqs.~(\ref{heat_additive}) and (\ref{Shannon_additive}), the total entropy production is decomposed as
\begin{equation}
\Sigma (XY) = \Sigma (X) + \Sigma (Y) - \Delta I(X:Y),
\label{entropy_decomposition}
\end{equation}
where  the total entropy production is not additive too, because of the mutual information term.  This observation is crucial to understand the consistency between Maxwell's demon and the second law, as discussed below.
We emphasize that the second law of thermodynamics always applies to the total entropy production:
\begin{equation}
\Sigma (XY) \geq 0.
\end{equation}
Correspondingly, a process is thermodynamically reversible if and only if $\Sigma (XY) = 0$.

We note that the terminology of the entropy ``production'' for the subsystems (i.e., $\Sigma (X)$ and $\Sigma (Y)$) is a little bit of an abuse.
More precisely, $\Sigma (X)$ is the sum of the entropy increase in $X$ and that in the bath associated with dynamics of $X$.
Strictly speaking, the terminology of ``production'' should be reserved for the entropy increase of the total system, not for that of a subsystem.
In the following, however, for the sake of simplicity, we refer to $\Sigma (X)$ and $\Sigma (Y)$ just as the entropy production of the subsystems.

\

$\Diamond\Diamond\Diamond$

\

\begin{figure}[htbp]
 \begin{center}
 \includegraphics[width=50mm]{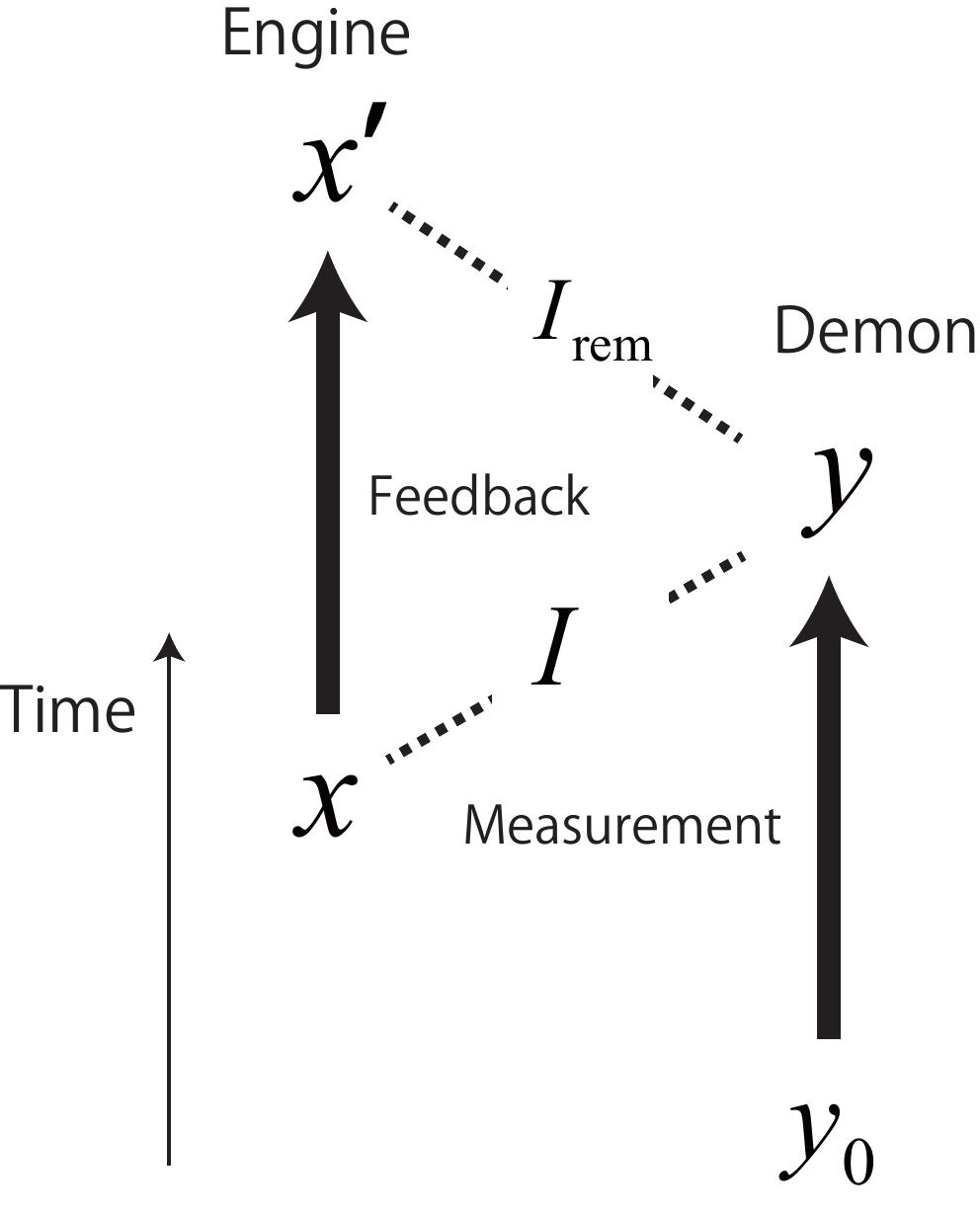}
 \end{center}
 \caption{Schematic of the measurement and the feedback processes, where $x, x'$ ($y_0, y$) represent the initial and the final states of the engine (the memory of the demon).  The initial correlation between the engine and the memory is assumed to be zero.  After the measurement of the engine by the demon, a correlation is established, which is represented by the mutual information $I$.  Feedback control is performed by using the measurement outcome $y$, and the remaining correlation after feedback is $I_{\rm rem}$. } 
 \label{Demon_entropy}
\end{figure}

\begin{figure}[htbp]
 \begin{center}
 \includegraphics[width=120mm]{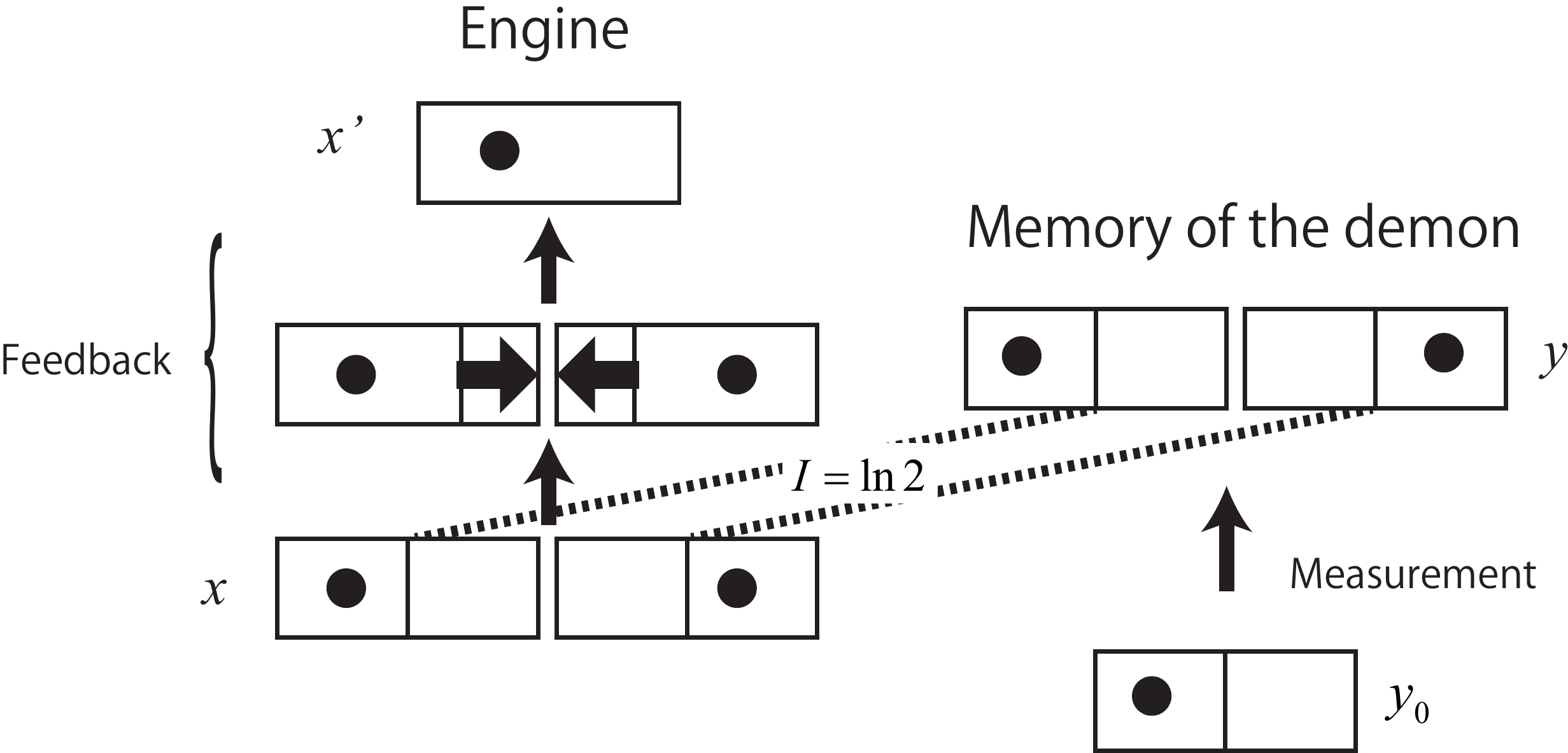}
 \end{center}
 \caption{A schematic of the Szilard engine and the memory of the demon, which is a special case of Fig.~~\ref{Demon_entropy}.
Here, both of the engine and the memory are represented by the two boxes with a particle.
} 
\label{Szilard_engine}
\end{figure}

We now consider stochastic thermodynamics of the measurement and feedback processes (see Fig.~\ref{Demon_entropy} for a schematic).
We suppose that subsystem $X$ is an engine measured and controlled by the demon, and subsystem $Y$ plays the role of the memory of the demon.
The Szilard engine discussed above is a special case of this setup; Fig.~\ref{Szilard_engine} shows the dynamics of the Szilard engine along with the memory of the demon.
To avoid too much complication, we do not explicitly formulate the computational states of the demon in the following, while it is straightforward to consider them~\cite{Sagawa2014}. 

We first consider the measurement process. Before the measurement, the system and the demon are not correlated, and the mutual information is zero.  Let $x$ be the initial state of the engine and $y_0$ the initial state of the demon.
During the measurement, the dynamics of the demon depends on the initial state $x$ of the system.  For simplicity, we assume that the measurement reads out the instantaneous value of $x$, and the system does not evolve during the measurement. After the measurement, the state of the demon, denoted as $y$, is correlated with $x$.   Here, $y$ is supposed to be equivalent to  the measurement outcome in Sec.~\ref{Sec_Work}.

Let $I$ be the mutual information between $x$ and $y$. The mutual-information change during the measurement is given by
\begin{equation}
\Delta I_{\rm meas} = I,
\end{equation}
which is positive if the demon gains information.
From Eq.~(\ref{entropy_decomposition}), the entropy production in the total system during the measurement is given by
\begin{equation}
\Sigma (XY)_{\rm meas} = \Sigma (X)_{\rm meas} + \Sigma (Y)_{\rm meas} -I.
\end{equation}
From the assumption that the system does not evolve during the measurement, $\Sigma (X)_{\rm meas} = 0$.  Therefore, we obtain
\begin{equation}
\Sigma (XY)_{\rm meas} = \Sigma (Y)_{\rm meas} -I.
\end{equation}
Since the second law applies to the total entropy production, $\Sigma  (XY)_{\rm meas} \geq 0$, we obtain
\begin{equation}
\Sigma (Y)_{\rm meas}  \geq I.
\label{second_measurement}
\end{equation}
This implies that the entropy production of the memory during the measurement is bounded from below by the mutual information.

In terms of the nonequilibrium free energy (\ref{noneq_free}), we rewrite inequality~(\ref{second_measurement}) as
\begin{equation}
W_{\rm meas} \geq \Delta F(Y)_{\rm meas} + TI,
\label{work_measurement}
\end{equation}
where $\Delta F(Y)_{\rm meas} := \Delta E(Y)_{\rm meas} - T \Delta S(Y)_{\rm meas}$.
Inequality~(\ref{work_measurement}) reveals the fundamental lower bound of the energy cost for the measurement.  Here, $TI$ on the right-hand side comes from the right-hand side of Eq.~(\ref{second_measurement}), and  represents the additional energy cost to obtain the mutual information $I$.  This inequality has been derived in Refs.~\cite{Sagawa-Ueda2009,Sagawa-Ueda2012b}.

We next consider the feedback process, where the dynamics of the engine depends on the measurement outcome $y$.  For simplicity, we assume that the memory does not evolve during the measurement (i.e., $y$ remains unchanged).  After the feedback, the final state of the system is $x'$, and the remaining correlation between $x'$ and $y$ is denoted as $I_{\rm rem}$.
The mutual-information change during feedback is then given by
\begin{equation}
\Delta I_{\rm fb} = I_{\rm rem} - I.
\end{equation}
This  is negative if the obtained information is used during the feedback  by the demon as discussed in Sec.~\ref{Sec_Work}.
We note that $I$, $I_{\rm rem}$, and  $\Delta I_{\rm fb}$ respectively equal $I(X:Y)$, $I(X':Y)$, and $\Delta I$ in the notations of Sec.~\ref{Sec_Work}.

From Eq.~(\ref{entropy_decomposition}), the entropy production in the total system during the feedback is given by
\begin{equation}
\Sigma (XY)_{\rm fb} = \Sigma (X)_{\rm fb} + \Sigma (Y)_{\rm fb} +I - I_{\rm rem}.
\end{equation}
From the assumption that the memory  does not evolve during the feedback, $\Sigma (Y)_{\rm fb} = 0$.  Therefore, we obtain
\begin{equation}
\Sigma (XY)_{\rm fb} = \Sigma (X)_{\rm fb} +I - I_{\rm rem}.
\end{equation}
Again since the second law applies to the total entropy production, $\Sigma  (XY)_{\rm fb} \geq 0$, we obtain
\begin{equation}
\Sigma (X)_{\rm fb}   \geq  - (I - I_{\rm rem} ).
\label{second_feedback}
\end{equation}
This implies that the entropy production of the system  during the feedback can be negative up to the minus of the used information by the feedback.
We note that inequality~(\ref{second_feedback}) is equivalent to inequality~(\ref{second_law_feedback2}) in Sec.~\ref{Sec_Work}, where $\Sigma (X)_{\rm fb}$  equals $\Delta S(X) - \beta Q$ in the notation of Sec.~\ref{Sec_Work}.
In the case of the Szilard engine, $\Sigma (X)_{\rm fb} =  -\ln 2$.
Such reduction of the entropy is the bare essential of the role of Maxwell's demon.

We note that thermodynamic reversibility is achieved if and only if, for the measurement and the feedback processes,
\begin{equation}
\Sigma (XY)_{\rm meas} = 0, \ \ \Sigma (XY)_{\rm fb} =0, 
\end{equation}
respectively.  A model of Maxwell's demon that satisfies both of these reversibility conditions has been proposed in Ref.~\cite{Horowitz2013}.

As a side remark, we consider information erasure from the memory after the feedback process.
In the erasure process, the memory does not interact with the engine, but solely goes back to the initial distribution only in contact with the heat bath.  In this process, the second law is given by
\begin{equation}
\Sigma (Y)_{\rm erase} := \Delta S(Y)_{\rm erase} - \beta Q_{\rm erase} \geq 0,
\label{second_erase1}
\end{equation}
which is nothing but the (generalized) Landauer principle (\ref{noneq_work_memory}).  In the quasi-static limit, we have $\Sigma (Y)_{\rm erase} = 0$. In terms of the work and the nonequilibrium free energy, inequality~(\ref{second_erase1}) is rewritten as
\begin{equation}
W_{\rm erase} \geq \Delta F(Y)_{\rm erase}.
\label{second_erase2}
\end{equation}

We assume the complete information erasure, in which after the information erasure, the probability distribution and the Hamiltonian of the memory completely return to the initial ones before the measurement.  
This assumption is satisfied if the memory is in the standard computational state with local equilibrium, before the measurement and after the erasure.
In this case, $\Delta F(Y)_{\rm meas} = - \Delta F(Y)_{\rm erase}$.  Therefore, by summing up inequalities (\ref{work_measurement}) and (\ref{second_erase2}), we obtain~\cite{Sagawa-Ueda2009}
\begin{equation}
W_{\rm meas} + W_{\rm erase} \geq TI.
\label{work_tradeoff}
\end{equation}

Inequality (\ref{work_tradeoff}) is the trade-off relation between the work for the measurement and that for the erasure, and sets the fundamental lower bound of the energy cost required for the memory. 
We remark that the lower bound of (\ref{work_tradeoff}) is given only by the mutual information, but does not depend on the details of the memory (e.g.,  symmetric or asymmetric).
This mutual-information term exactly compensates for the additionally extractable work by feedback control (i.e., the mutual-information term in inequality (\ref{second_law_feedback6})).

\

$\Diamond\Diamond\Diamond$

\

We now summarize the key observation in the foregoing argument.
First of all, the measurement and feedback processes are \textit{individually} consistent with the second law, because $\Sigma (XY) \geq 0$ holds for the individual processes.
In this respect, there is not any contradiction between the second law and Maxwell's demon.

The apparent ``paradox'' of Maxwell's demon would stem from the negative entropy production of the engine, $\Sigma (X)_{\rm fb} < 0$.
However, the second law must apply to the total system, and therefore the negative entropy production of the subsystem is not  a contradiction.  If we take into account the change in the mutual information, by adding it to $\Sigma (X)_{\rm fb}$ as $\Sigma (X)_{\rm fb} + (I - I_{\rm rem})$, we recover the total entropy production $\Sigma (XY)_{\rm fb}$ that is always nonnegative.

In the case of the Szilard engine, $\Sigma (X)_{\rm fb}= - \ln 2$  and $I - I_{\rm rem} = \ln 2$.
Therefore, the total entropy production is just zero: $\Sigma (XY)_{\rm fb} = - \ln 2 + \ln 2 = 0$, which  implies that the Szilard engine is a reversible information engine.
Table~\ref{Table_entropy_balance_Szilard} summarizes the entropy balance of the Szilard engine for the case that the measurement, the feedback, and the erasure processes are all quasi-static.

\begin{table}
\caption{The entropy balance of the Szilard engine, where $X$ is the engine and $Y$ is the demon. Here, we assumed that all the processes (i.e., measurement, feedback, and erasure) are quasi-static.}
\begin{center}
\begin{tabular}{| l || l | l | l | l | }\hline
{} & $\Sigma (XY)$ & $\Sigma (X)$ & $\Sigma (Y)$ & $\Delta I$ \\ \hline \hline
Measurement & $0$ & $0$ & $\ln 2$ & $\ln 2$ \\ \hline
Feedback & $0$ & $-\ln 2$ & $0$ & $-\ln 2$ \\ \hline
Erasure & $0$ & $0$ & $0$ & $0$  \\ \hline
\end{tabular}
\end{center}
\label{Table_entropy_balance_Szilard}
\end{table}

As discussed above, an information-erasure process can follow the feedback process.
We emphasize that, however, we do not necessarily need to consider information erasure to understand the consistency between the demon and the second law.

\section{Concluding remarks}
\label{Sec_Concluding}

In this article, we have only focused on the second law of thermodynamics at the level of the ensemble average, with which we have clarified the concept of reversibilities and the entropy production.
However, stochastic thermodynamics has much richer aspects, which we did not discuss so far.
In the following, we will briefly summarize some important topics beyond the scope of this article.

\paragraph{Fluctuation theorem.}
One of the most important discovery in stochastic thermodynamics is the fluctuation theorem~\cite{Cohen1993,Gallavotti,Evans2002,Jarzynski1997,Crooks1999,Jarzynski2000,Seifert2005}.  Roughly speaking, we consider the stochastic version of the entropy production $\sigma$, which gives  $\Sigma = \langle \sigma \rangle$ with $\langle \cdots \rangle$ being the ensemble average.  Then,  the fluctuation theorem (or more precisely, the integral fluctuation theorem or the Jarzynski equality) is given by
\begin{equation}
\langle e^{-\sigma} \rangle = 1,
\label{FT}
\end{equation}
which implies that the second law of thermodynamics can be represented by an equality, if we take into account fluctuations of the entropy production.  By using the convexity of the exponential function, we have $\langle e^{-\sigma} \rangle \geq e^{-\langle \sigma \rangle}$.  Therefore, Eq.~(\ref{FT}) reproduces the usual second law $\langle \sigma \rangle \geq 0$. 
We note that the fluctuation-dissipation theorem and its generalization to nonlinear responses can be obtained from the fluctuation theorem~(\ref{FT})~\cite{Andrieux2007,Saito2008}.

Thermodynamics of information can be formulated at the level of the stochastic entropy production, and thus the fluctuation theorem can be generalized by incorporating the mutual information~\cite{Sagawa-Ueda2010,Sagawa-Ueda2012b,Sagawa-Ueda2013}.

\paragraph{Autonomous demons.}
In Secs.~\ref{Sec_Work} and \ref{Sec_Entropy}, we have discussed Maxwell's demon that performs a single measurement-feedback process.
We can extend  the second law and the fluctuation theorem to multiple measurement-feedback processes~\cite{Horowitz2010,Sagawa-Ueda2012a}, and further to situations that measurement and feedback are performed autonomously and continuously in time~\cite{Allahverdyan2009,Suzuki,Strasberg2013,Ito2013,Hartich2014,Munakata2014,Horowitz2014a,Horowitz2014b,Shiraishi2015a,Shiraishi2015b,Yamamoto2016,Rosinberg2016,Hartich2016}.  Here, the informational quantities that characterize continuous information flow, such as the transfer entropy~\cite{Schreiber2000} and the learning rate (or just the ``information flow'')~\cite{Horowitz2014a,Hartich2016},  play crucial roles. 

There is also another formulation of autonomous demons based on the concept of information reservoirs~\cite{Mandal2012,Deffner2013,Barato2014b,Barato2014c,Merhav2015,Boyd2016b}.
These two approaches (the autonomous measurement-feedback approach and the information reservoir approach) are  shown equivalent in general~\cite{Strasberg2017}.
It has also been shown that there is an exact mapping between these approaches for a typical model~\cite{Shiraishi2016b}, based on the concept of partial entropy production~\cite{Shiraishi2015a}.
We note that other informational quantities, such as the Kolmogorov-Sinai entropy, have been investigated in the context of thermodynamics~\cite{Boyd2016a,Boyd2017}.

\paragraph{Application to biological systems.}
Interesting applications of thermodynamics of information, especially the theory of autonomous demons, are also found in biophysics.  In fact,  living cells perform autonomous information processing based on biochemical reactions;
Thermodynamics of information in biochemical systems is now an active emerging field~\cite{Lan2012,Mehta2012,Barato2013,Barato2014a,Lang2014,Sartori2014,Ito2015,Kobayashi2015,Sartori2015,Ouldridge2017a,Ouldridge2017b,Matsumoto2017}.

\paragraph{Quantum thermodynamics and quantum information.}
We have focused on classical thermodynamics and classical information so far, while stochastic thermodynamics also applies to quantum systems~\cite{Kurchan2000,Tasaki2000,Esposito2009,Campisi2011,Sagawa2012,Alhambra2016,Strasberg2017}.
Quantum analogues of the Szilard engine have been proposed~\cite{Zurek1986,SWKim2011}, and the role of quantum information in thermodynamics has been intensively investigated~\cite{Lloyd1989,Nielsen1998,Sagawa-Ueda2008,Sagawa-Ueda2009,Morikuni2011,Vedral2011,Sagawa2012a,Funo2013,Tajima,Park2013,Goold}.  Furthermore, several experiments on thermodynamics of information have been performed in the quantum regime~\cite{Camati2016,Cottet2017,Masuyama2017}.  We also note that there is another interesting approach to quantum thermodynamics, called thermodynamic resource theory~\cite{Horodecki2013,Brandao2015}.

\paragraph{Ultimate origin of the information-thermodynamics link.}
Last but not least, the fundamental origin of the information-thermodynamics link is yet to be fully understood based on quantum mechanics.
Throughout this manuscript, we have assumed that there exists a large heat bath in thermal equilibrium, specifically in the canonical distribution. 
However, the microscopic characterization of thermal equilibrium is quite nontrivial, because a typical pure quantum state~\cite{Popescu2006} and even a single energy eigenstate~\cite{Rigol2008} can behave as thermal.
In this context, the eigenstate-thermalization hypothesis (ETH) has been considered to be a plausible mechanism of thermalization in isolated quantum systems~\cite{Rigol2008}.  Based on the ETH, the second law and the fluctuation theorem have been proved in the short time regime for isolated quantum many-body systems where the heat bath is initially in a single energy eigenstate~\cite{Iyoda2017}. 

\

$\Diamond\Diamond\Diamond$

\

In these decades, there has been significant progress in stochastic  thermodynamics, which has led to the modern theory of thermodynamics of information.
Stochastic thermodynamics is still quite a hot field, and thermodynamics of information would further lead to the fundamental understanding of the interplay between physics and information.

\paragraph{Acknowledgments.}
The author is grateful to Masahito Ueda, John Bechhoefer, Jordan M. Horowitz, and Naoto Shiraishi  for a lot of valuable comments, and to Christian Van den Broeck, Massimiliano Esposito, and Udo Seifert for valuable suggestions. 
This work is supported by JSPS KAKENHI Grant No. JP16H02211 and No. JP25103003.


\end{document}